\documentclass[useAMS]{mn2e}
\usepackage{graphicx}
\usepackage{amssymb}
\usepackage{lscape}

\title[Deep GMRT 150 MHz observations of LBDS field]{Deep GMRT 150 MHz
observations of the LBDS-Lynx region: Ultra-Steep Spectrum Radio Sources}
\author[C. H. Ishwara-Chandra et al.]{C. H. Ishwara-Chandra$^{1,2}$\thanks{E-mail:
ishwar@ncra.tifr.res.in}, S. K. Sirothia$^1$\thanks{E-mail:sirothia@ncra.tifr.res.in}, Y. Wadadekar$^1$\thanks{E-mail:yogesh@ncra.tifr.res.in}, S. Pal$^3$\thanks{E-mail:spal@cyllene.uwa.edu.au}, R. Windhorst$^4$\thanks{E-mail:Rogier.Windhorst@asu.edu}\\
$^1$National Centre for Radio Astrophysics, TIFR, Post Bag No. 3., Ganeshkhind, Pune 411007, India\\
$^2$Inter-University Centre for Astronomy and Astrophysics, Post Bag 4, Ganeshkhind, Pune 411007, India\\
$^3$International Centre for Radio Astronomical Research, University of Western Australia, 35 Stirling Highway, WA, 6009, Australia\\
$^4$School of Earth \& Space Exploration, 
Arizona State University P.O. Box 871404 Tempe, AZ 85287-1404, USA}

\begin{document}

\date{}

\pagerange{\pageref{firstpage}--\pageref{lastpage}} \pubyear{2009}

\maketitle

\label{firstpage}

\begin{abstract}

It has been known for nearly three decades that high redshift radio
galaxies exhibit steep radio spectra, and hence ultra-steep spectrum
radio sources provide candidates for high-redshift radio galaxies. Nearly
all radio galaxies with z $>$ 3 have been found using this
redshift-spectral index correlation. We have started a programme with
the Giant Metrewave Radio Telescope (GMRT) to exploit this correlation
at flux density levels about 10 to 100 times deeper than the known high-redshift
radio galaxies which were identified primarily using the already available
radio catalogues.  In our programme, we have obtained deep, high
resolution radio observations at 150 MHz with GMRT for several $'${\it deep}$'$ fields
which are well studied at higher radio frequencies and in other bands
of the electromagnetic spectrum, with an aim to detect candidate high
redshift radio galaxies.  In this paper we present results from the
deep 150 MHz observations of LBDS-Lynx field, which has been already imaged
at 327, 610 and 1412 MHz with the Westerbork Synthesis Radio Telescope (WSRT) and at 1400 and 4860 MHz with
the Very Large Array (VLA). The 150 MHz image made with GMRT has a rms noise of $\sim$
0.7 mJy/beam and a resolution of $\sim 19^{''} \times 15^{''}$. It is
the deepest low frequency image of the LBDS-Lynx field. The source
catalog of this field at 150 MHz has about 765 sources down to $\sim$
20\% of the primary beam response, covering an area of about 15
degree$^2$.  Spectral index was estimated by cross correlating each
source detected at 150 MHz with the available observations
at 327, 610, 1400 and 4860 MHz and also using available radio surveys such as 
WENSS at 327 MHz and
NVSS and FIRST  at 1400 MHz. We find about 150 radio sources with
spectra steeper than 1. About two-third of these are not detected in
Sloan Digital Sky Survey, hence are strong candidate high-redshift radio galaxies, which 
need to be further explored  with deep infra-red imaging and
spectroscopy to estimate the redshift.
\end{abstract}

\begin{keywords}
galaxies: active -  galaxies: high-redshift - radio continuum: galaxies 
\end{keywords}

\section{Introduction}

Since powerful radio sources reside in massive ellipticals (Best et
al. 1998), finding radio galaxies at high redshifts is important to
understand the radio evolution of galaxies.  It had been
noticed in the early 80's that the fraction of radio sources that can
be optically identified is lower by a factor of 3 or more for steep
spectrum radio sources ($\alpha > 1; S_\nu \propto \nu^{-\alpha}$; 
Tielens, Miley \& Wills, 1979, Blumenthal \& Miley, 1979; Gopal-Krishna \& Steppe, 1981),
suggesting that the radio sources with steeper spectra at decimeter wavelengths 
are more distant compared to the ones with normal spectra ($ 1 > \alpha >
0.5$). Over the years, this correlation has been exploited to search
for radio sources at high redshifts (Rottgering et al., 1994; De Breuck et al. 2000,
2002; Klamer et al. 2006). Since the radio emission does not suffer from dust absorption,
selecting candidate high redshift radio galaxies (HzRGs) at radio
frequencies provides an optically unbiased sample. The most distant
radio galaxy at z = 5.19 was discovered using the spectral index -
redshift correlation (van Breugel et al. 1999). Until today, about 45
radio galaxies are known beyond redshift of 3, and nearly all of
them were discovered through the radio spectral index - redshift
correlation. Thus, this is the most efficient method to find HzRGs.

In addition to constraining models of formation and evolution of
massive galaxies, HzRGs can also be used to study the environment
at those epochs. Deep radio polarimetric observations of several HzRGs
have shown large rotation measures (RM) of the order of thousands of
rad m$^{-2}$ (Pentericci et al. 2000; Carilli et al. 1997; Athreya et
al. 1998; Kronberg et al. 2008).  Pentericci et al. (2000) also found that the fraction of
radio galaxies with extreme Faraday rotation is increasing with
redshift which is consistent with the hypothesis that the environment
is denser at high redshift. Since at moderate redshifts large RMs are known to occur
in radio
galaxies residing near the center of clusters of galaxies,
the HzRGs are an excellent tool to study the (proto) cluster
environment at high redshifts. It is therefore important to identify
and study as many HzRGs as possible.

A major programme to search for HzRGs using the radio spectral index -
redshift correlation was carried out in the last decade using
different radio surveys at 1400 MHz and lower frequencies (eg:
Rottgering et al. 1994, De Breuck et al. 2000, 2002, 2004, 2006; Klamer et al. 2006; see also review 
by Miley and De
Breuck, 2008).  The searches were mostly limited by the sensitivity
limit of these shallow, wide-area radio surveys. This bias has led to the
detection of only the brightest HzRGs which are at the top end of
the radio luminosity function (See section 2), which is about three
orders of magnitude brighter than the luminosities at the lower end of
FR-II population (Fanaroff \& Riley, 1974). Since  steep spectrum
radio sources are preferentially selected at low radio frequencies,
deeper observations at these frequencies are needed to discover HzRGs
that are 10 to 100 times less luminous than most of the known HzRGs, and these
could be the high-redshift counterparts of more typical FR-II radio
galaxies. The 150 MHz band of GMRT (Giant Metrewave Radio Telescope,
India, http://www.ncra.tifr.res.in) with its large field of view
(half-power beam width of 3 degrees), high angular resolution ($\sim
20^{''}$) and better sensitivity ($\sim$ 1 mJy from a full synthesis
observation) is well suited for this kind of work. 
For example, a steep spectrum radio source ($\alpha > 1$) 
at the completeness limit of WENSS at
327 MHz (Rengelink et al. 1997), will have a flux density $>$ 65 mJy
at 150 MHz.  Our experience with GMRT at 150 MHz shows that it is
possible to reliably detect sources more than 10 times fainter than this
value.
The large field of view of GMRT also enables us to detect close to
thousand radio sources in a single pointing.  
Encouraged by this, we have started a programme to
observe several carefully chosen deep fields at 150 MHz with GMRT with
an aim to detect steep spectrum radio sources to flux density levels much
fainter than that of known high-redshift radio sources (see Section 2).
As the sample at low radio frequencies goes deeper, the
available surveys at higher radio frequencies such as FIRST
won't be deep enough to get the spectral index estimates, particularly for 
steep spectrum sources. For example,
from the present work, it is seen that if we use only the FIRST catalogue,
37\% of the sources below 18 mJy at 150 MHz (which corresponds to 1 mJy at 
1.4 GHz for a spectral index of 1.3) do not have counter parts in FIRST
as against 9\% for stronger sources. This limitation can be addressed
by deeper surveys of this region at higher radio frequencies.
Therefore, we have chosen the well known deep fields because significant amount of
data already exist for them at higher radio frequencies and in the optical and
infrared bands. These deep observations will allow us to estimate the 
spectral index for faint sources which are below the detection limit
of NVSS or FIRST. The deep optical data will help to estimate the redshifts
and eliminate nearby and known objects among the steep spectrum sources. 
Wherever needed, additional deep observations with the
GMRT at 610 MHz will be obtained for a better estimate of radio
spectra. In addition to the steep spectrum radio sources, this kind of
survey will also help us to understand the evolution (LogN-LogS) and
spectral index properties of faint radio sources, to detect low-power
FRI sources, relic emission and large angular size radio sources.

In this paper we present deep 150 MHz radio observations of the
LBDS-Lynx field with the GMRT with the primary aim of detecting steep
spectrum radio sources which are candidate HzRGs.  The Leiden-Berkeley
Deep Survey (LBDS) in the Lynx area (Windhorst et al. 1984) was
carried out in the mid eighties to better understand the nature of
faint radio sources and their cosmological evolution. Multi-band
optical data were obtained with the Kitt Peak Mayall 4m telescope and
subsequently deep radio observations were carried out at 327 MHz and
1412 MHz with WSRT, complimented by 1400 MHz and 4860 MHz imaging with the VLA
(Windhorst et al 1984, 1985, Oort, 1987, Oort et al. 1988).  These
radio observations showed a moderate increase in the number of radio
sources below $\sim$ 5 mJy at 1.4 GHz, which were predominantly
identified with blue galaxies, a population responsible for the upturn in the
source counts at faint flux density levels (Windhorst et al. 1985).  The
spectral index studies of radio sources selected at 327 MHz in this region using the 327 MHz and
1400 MHz data showed that the median spectral index was flatter for
fainter sources selected at 327 MHz (Oort et al. 1988).  The sources
with the steepest spectra ($\alpha > 1$) were mostly unidentified in
the optical 4m telescope images, once again indicating that they were
likely to be very distant.

The sources we detected at 150 MHz with the GMRT were compared with
the deep observations of this region at 327, 610 and 1412 MHz with
the WSRT and at 4860 MHz with the VLA. We have also used the available
catalogues such as WENSS at 327 MHz and NVSS and FIRST at 1400 MHz to
estimate spectral indices.
We identify about 150 steep spectrum radio
sources from this field. About two-third of them are not detected to
the limit of SDSS, hence are strong candidate HzRGs. In Section 2,
we present supporting arguments for reviving the search for HzRGs using
the steep spectrum technique.  Observations, data analysis and the
data obtained are presented in Section 3. Results and discussions are
presented in Section 4 and concluding remarks in Section 5.

\section{Re-visit of Ultra-steep spectrum technique for finding
HzRGs}

The method to find HzRGs using the correlation between steep radio spectrum 
and high redshift of the radio galaxy is the most efficient technique 
to discover HzRGs (e.g: Pedani, 2003; de Breuck et al. 2000). Till date
about 45 radio galaxies have been discovered
with z $>$ 3 using this technique. Miley \& de Breuck (2008) in their
review have provided a table of all known HzRGs with redshift more than 2.
We have independently surveyed the literature for an update on all HzRGs
with redshift more than 3, which is provided in Table 1. A total of 47 galaxies are listed in the table,
which is arranged as follows: Column 1: Source Name; 
Column 2: redshift; Column 3: radio spectral index (in a few cases where 
the spectral index was not known, we have assumed a value of 1); Column 4: observed
flux density at 1.4 GHz; Column 5: estimated flux density at 150 MHz using the spectral 
index and the 1.4 GHz flux density, given in Columns 3 and 4; Column 6: Reference.

\begin{table}
\caption{Known HzRGs till date, discovered mostly using ultra steep spectrum technique.}
{\footnotesize
\begin{tabular}{l l l l l l }
Source Name        &  z    & $\alpha$ &S$_{1400}$&S$_{150}$& Ref\\
                   &       &       & mJy  & mJy   &    \\
                   &       &       &        &         &    \\
B30032+412        &3.658 &1.16 & 102  & 1330 & Jar01\\
NVSSJ012142+132058&3.516 &1.37 &  55.4  & 1150 & deB01\\
B20140+32         &4.413 &1.17 &  91.5  & 1220 & Raw91\\
NVSSJ020510+224250&3.5061&1.37 &  58.4  & 1212 & deB00\\
NVSSJ021308-322338&3.976 &0.90 &  30.4  & 223 & deB06\\
NVSSJ023111+360027&3.079 &1.30 &  44.3  & 788 & deB00\\
PMNJ0253-2708     &3.16  &1.03 & 273.3& 2673 &    McC96\\
RCJ0311+0507      &4.514 &1.31 & 473.0 & 8597    &    Kop06\\
PKS0315-257       &3.1307&1.05 & 424.0 & 4334 &    McC90\\
NVSSJ033953-232136&3.49  &1.00$^*$ &  36.5  & 334     &    Cac00\\
NVSSJ034642+303949&3.72  &1.42 &  34.9  & 809 &    deB02\\
4C+60.07          &3.791 &1.45 &  156.8 & 3885 &    Cha96\\
WNJ0617+5012      &3.153 &1.37 &  26.5  & 550 &    deB00\\
4C+41.17          &3.792 &1.51 &  235   & 6650 &    Cha96\\
WNJ0747+3654      &2.992 &1.41 &  36.8  & 835   &    deB00\\
NVSSJ075806+501104&2.996 &1.07 &  98.0  & 1047 &    Cru06\\
NVSSJ083609+543325&3.341 &1.02 &  62.7  & 600 &    Cru07\\
B20902+34         &3.382 &0.84 &  313   & 2010 &    Lil88\\
TNJ0924-2201      &5.19  &1.63 &  73.3  & 2705 &    van99\\
NVSSJ094724-210505&3.377 &1.00$^*$ &  10.9  & 100     &    Bro06\\
NVSSJ095438-210425&3.431 &1.00$^*$ &  51.3  & 469     &    Bro06\\
NVSSJ095751-213321&3.126 &1.00$^*$ &  62.3  & 570     &    Bro06\\
NVSSJ104906-125819&3.697 &1.51 &  108.1 & 3059 &    Bor07\\
6CB105034+544035  &3.083 &1.34 &  66.4  & 1290 &    deB00\\
NVSSJ105917-303658&3.263 &1.06 &  64.7  & 676   &    Bry09\\
NVSSJ111223-294807&3.09  &1.40 &  101.0 & 2240 &    deB00\\
B21121+31B        &3.2174&1.46 &  74.0  & 1875 &    deB00\\
TNJ1123-2154      &4.109 &1.57 &  49.6  & 1603    &    deB00\\
4C+39.37          &3.22  &1.47 &  255.3 & 6612 &    Raw90\\
4C+03.24          &3.5699&1.28 &  256   & 4354  &    Rot97\\
8C1435+635        &4.261 &1.31 &  451   & 8197    &    Lac94\\
TNJ1338-1942      &4.11  &1.31 &  122.9 & 2234 &    deB00\\
J163912+405236    &4.88  &0.75 &  21.8  & 114.7 & Jar09\\
WN B1702+6042     &3.223 &1.24 &  53.2  & 828     &    Vil99\\
7C1814+670        &4.05  &1.09 &  236.1 & 2637    &    Lac99\\
4C+72.26          &3.536 &1.22 &  259   & 3857 &    deB01\\
TXS1911+636       &3.59  &1.42 &  23.3  & 540 &    deB01\\
MRC2005-134       &3.837 &1.42 &  115.4 & 2676 &    deB01\\
TNJ2009-3040      &3.158 &1.36 &  65.3  & 1326    &    Bor07\\
4C+18.61          &3.056 &0.87 &  575.3 &  3948   &    Ste99\\
4C+19.71          &3.594 &1.02 &  306   & 2927 &    Ste99\\
NVSSJ230123-364656&3.22  &1.38 &  20.3  & 431 &    deB06\\
WNJ2313+4053      &2.99  &1.50 &  11.3  & 313   &    deB00\\
NVSSJ231402-372925&3.45  &1.22 &  129.9 & 1934 &    deB06\\
NVSSJ231727-352606&3.874 &1.19 &  59.2  & 825 &    deB06\\
NVSSJ232100-360223&3.32  &1.65 &  15.1  & 583 &    deB06\\
B3J2330+3927      &3.087 &1.21 &  148.4 & 2162 &    Per05\\
\end{tabular}
}
\\
$^*$The radio spectral index measurements were not available; so a spectral
index of 1 was assumed. \\ References: 
{\sf Jar01:} Jarvis et al. 2001;
{\sf deB01:} de Breuck et al. 2001;
{\sf Raw91:} Rawlings et al. 1991;
{\sf deB00:} de Breuck et al. 2000;
{\sf deB06:} de Breuck et al. 2006;
{\sf McC96:} McCarthy et al. 1996;
{\sf Kop06:} Kopylov et al. 2006;
{\sf McC90:} McCarthy et al. 1990;
{\sf Cac00:} Caccianiga et al, 2000;
{\sf deB02:} de Breuck et al. 2002;
{\sf Cha96:} Chambers et al. 1996;
{\sf Cru06:} Cruz et al. 2006;
{\sf Cru07:} Cruz et al. 2007;
{\sf Lil88:} Lilly, 1988;
{\sf van99:} van Breugel et al. 1999;
{\sf Bro06:} Brookes et al. 2006;
{\sf Bor07:} Bornancini et al. 2007;
{\sf Bry09:} Bryant et al. 2009;
{\sf Raw90:} Rawlings et al. 1990;
{\sf Rot97:} Rottgering et al. 1997;
{\sf Lac94:} Lacy et al. 1994;
{\sf Jar09:} Jarvis et al. 2009;
{\sf Vil99:} Vilani et al. 1999;
{\sf Lac99:} Lacy et al. 1999;
{\sf Ste99:} Stern et al. 1999;
{\sf deB06:} de Breuck et al. 2006;
{\sf Per05:} Perez-Torres et al. 2005.
\end{table}

In order to understand whether the known HzRGs represent typical FRII
radio sources at high-redshifts, or the highest luminosity sources in
that category, we have computed the expected flux density at 150 MHz
corresponding to the FRI/FRII break luminosity (assuming standard
cosmological parameters of $\Omega_m$ of 0.27, $\Omega_{vac}$ of 0.73
and H$_o$ of 71 kms$^{-1}$Mpc$^{-1}$; Bennett et al. 2003).  We assume here that the FRI/FRII
break luminosity does not evolve significantly with redshift.  At
a redshift of 3, the FRI/FRII break luminosity corresponds to a flux
density of $\sim$ 4 mJy at 150 MHz (taking a spectral index of 1 for
the K-correction). The corresponding values at redshift of 4 and 5 will be
$\sim$ 1.9 and $\sim$ 1.1 mJy, respectively.

\begin{figure}
\includegraphics[width=59mm,angle=270]{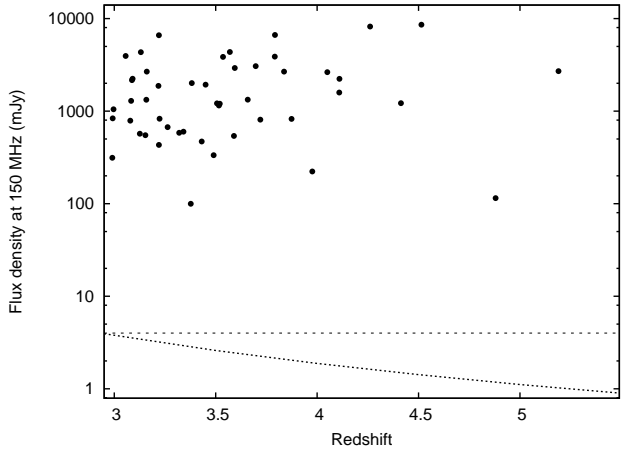}
\caption{The 150 MHz flux densities of known HzRGs, extrapolated using
the available spectral index and flux density measurements.  
The dotted line at the bottom indicates 
the observed 150 MHz flux density corresponding to the rest-frame FRI/FRII break
luminosity. The dashed horizontal line is the GMRT detection limit from the 
present 150 MHz observation. It is clear from this figure that a large number 
'normal population' of FRIIs, that are 10 to 100 times less luminous than the known HzRGs are
yet to be discovered.  }
\end{figure}

In Figure 1, we plot the observed flux densities at 150 MHz of all known
HzRGs, computed using the available flux density measurements and the spectral
index. The dotted line corresponds to the 150 MHz flux density at the
FRI/FRII break luminosity computed for different redshifts, assuming $\Lambda$
of 0.73 and $\Omega_M$ of 0.27.  It is clear from the figure that
nearly all of the known HzRGs are two to three orders of magnitude more
luminous than the FRI/FRII break luminosity.  This is largely due to
selection effects because nearly all the searches for steep spectrum
radio sources use sky surveys such as WENSS and SUMSS (Mauch et al. 2003)
 at low radio
frequencies and NVSS at higher radio frequencies. This selection is
biased towards stronger radio sources. The median flux density at 1400 MHz of the
known HzRGs is $\sim$ 70 mJy and the corresponding flux density at 150 MHz is
$\sim$ 1.3 Jy (90\% of them are stronger than $\sim$ 0.5 Jy), whereas
the FRI/FRII break luminosity is more than two orders of magnitude
fainter.

The above argument implies that the known HzRGs represent the tip of
the ice-berg in flux density.  There are, potentially, a large number
of HzRGs yet to be discovered which are expected to be 10 to 100 times less luminous
than the known HzRGs (see Fig. 1).  Although initially it was believed
that there is a 'cut-off' in the radio luminosity function at
high-redshift, subsequent to the discovery of several radio galaxies
at z $>$ 4, this evidence became suspect (Jarvis et al 2001). The radio luminosity
function (RLF) of
HzRGs indicate that at luminosities 10 to 100 times lower than the
known HzRGs, we could expect at least a 10 fold increase in space density
of the radio sources beyond the redshift of 3 (Waddington et al. 2001).

Systematic efforts are needed to find this population of radio
galaxies, which are not at the brightest end of the radio luminosity
function.  As mentioned earlier, the most efficient method to find
high-redshift radio galaxies is through the steep spectrum selection.
This large gap, can thus be filled by searching for steep spectrum sources
using deep radio observations at low frequencies.  The present
observation of the LBDS field at 150 MHz with the GMRT has an rms noise of
$\sim$ 0.7 mJy/beam, and the source catalog has about 750 sources
 having a flux density of $\gtrsim$ 4 mJy.  The sky coverage is
about 7 degree$^2$ for half-power beamwidth and 
15 degree$^2$ to the 20\% of the peak primary
beam response in a given pointing. A source at the threshold of detection in the FIRST survey
will have a flux density of $>$ 10 mJy at 150 MHz for a spectral index
steeper than 1.  Compared to the 150 MHz flux densities of known
HzRGs, this value is close to two orders of magnitude fainter. In the
present case of LBDS-Lynx field, deep observations exist at 327,
610 and 1400 MHz, which allowed us to obtain reliable spectral index
estimates for sources as faint as 7 mJy at 150 MHz.
Using our observations, we have obtained about 100 radio sources with
spectral index steeper than 1 having no optical counterparts in the
SDSS.  The median 150 MHz flux density for these sources is $\sim$ 110 mJy,
which is an order of magnitude fainter than the median flux density at 150 MHz
for known HzRGs.

\begin{figure}
\includegraphics[width=83mm,angle=0]{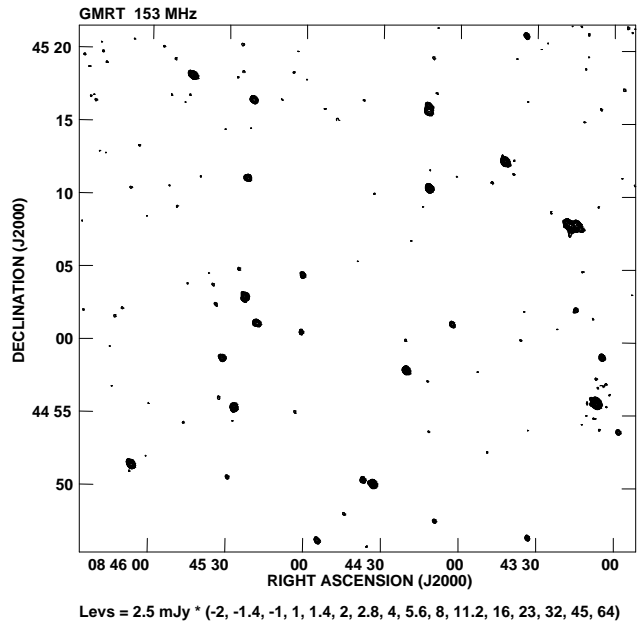}
\caption{A portion of the 150 MHz image of LBDS field obtained with the GMRT. 
The rms noise in the image is $\sim$ 0.7 mJy/beam and the resolution is
$\sim 19^{''} \times 15^{''}$ with position angle of 27$^\circ$. }
\end{figure}

\section{Observation and data analysis}

\subsection{GMRT 150 MHz observations}
The 150 MHz observations of the LBDS-Lynx field centered at
RA=08h41m46s and DEC=+44d46$'$50${''}$ (J2000) were carried out with the Giant
Metrewave Radio Telescope (GMRT) on December 11, 2006 using a
bandwidth of 16 MHz, between local midnight to morning in order to reduce the
radio frequency interference (RFI). The GMRT consists of 30
antennas, each of 45 meter diameter located 90 km from Pune, India and
operates at five frequency bands from 150 MHz to 1450 MHz (Swarup et
al. 1991).  For determining the flux density scale, standard flux
calibrators (3C147 and 3C286) were observed and their flux densities at
150 MHz was estimated using the Perley-Taylor scale (Perley \& Taylor 1991). From the estimate
of median spectral index of the sources in the field (0.78), which was
close to the expected value (e.g: Oort, Steemers \& Windhorst, 1988),
we infer that the error in the flux density scale
is $\lesssim$ 10\% at 150 MHz.  The data were recorded in the spectral
line mode with 256 channels of 62.5 kHz bandwidth per channel at 150
MHz to minimise the effect of bandwidth smearing.
A phase calibrator was observed for five minutes for every 30
minutes of on source observation.  The data were analysed using {\tt
AIPS++} (v1.9;build \#1556; see Sirothia 2008, Sirothia et al. 2009).  
While calibrating the data, bad data
points were flagged at various stages.  The data for antennas with
relatively large errors in antenna-based gain solutions were examined
and flagged over certain time ranges. Some baselines were flagged,
based on closure errors on the bandpass calibrator.  Channel and
time-based flagging of the data points corrupted by radio frequency
interference (RFI) was done by applying a median filter with a
$6\sigma$ threshold.  Residual errors above $5\sigma$ were also
flagged after a few rounds of imaging and self-calibration.
The final image was made after several rounds of phase self-calibration,
and one round of amplitude-and-phase self-calibration, where the data were normalized
by the median gain found for the entire data. The primary
beam correction was applied to the final image. The final rms
noise achieved at 150 MHz near the center of the field in 
a source free region 
was $\sim$ 0.7 mJy/beam with a resolution of $\sim 19^{''} \times 15^{''}$ at position
angle of 27$^\circ$. A portion of the image is presented in Figure 2.

\begin{figure}
\includegraphics[width=50mm,height=85mm,angle=270]{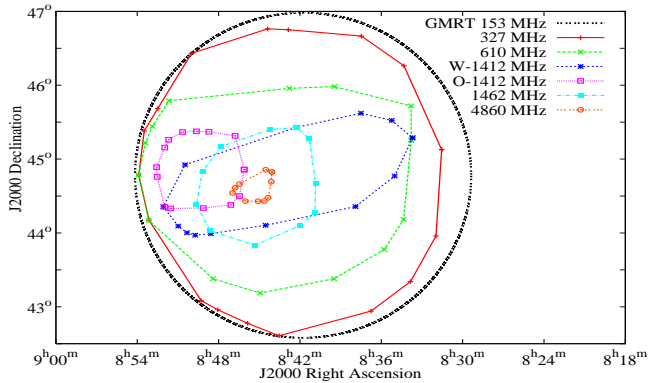}
\caption{Coverage of different existing deep observations of LBDS field
at 327, 610, 1412 and 1465 MHz with respect to the GMRT field of view.
One set of observations at 1412 MHz cover a  much smaller area but are 
deeper compared to another set of observation at 1412 MHz.
WENSS, NVSS and FIRST cover the full area of our 150 MHz coverage. They are  not indicated
in this figure (Also see Table 2)}
\end{figure}

\begin{table}
\caption{Summary of radio observations and number of radio sources
available in the LBDS-Lynx area, within
20\% of the primary beam of GMRT at 150 MHz (Figure 3).  }
\begin{tabular}{lllll}
     &      &        &       &            \\
Freq. & Telescope  &  1 $\sigma$ noise   & No. of  &  Ref.      \\
(MHz) &      &  (mJy) & Sources&            \\
     &      &        &       &            \\
153  & GMRT &  0.7   &  765  & This paper \\
327  & WSRT &  0.9   &  394  & Oort88 \\
327  & WSRT-WENSS &  4     &  302  & Reng97\\
610  & WSRT &  0.3   &  437  & WindPvt \\
1400 & VLA-FIRST &  0.2   &  1327 & Beck95\\
1400 & VLA-NVSS &  0.7   &  789  & Cond98\\
1412 & WSRT &  0.2   &  231  & Wind84\\
1412 & WSRT &  0.02   &  349  & Oort87\\
1462 & VLA &  0.045   &  139  & Wind85\\
4860 & VLA &  0.015   &  59   & Donn87\\
\end{tabular}
References: ~ Oort88: Oort et al. 1988, Reng97: Rengelink et al. 1997, 
WindPvt: Windhorst, unpublished data, Beck95: Becker et al. 1995,
Cond98: Condon et al.  1998, Wind84: Windhorst et al.  1984,
Oort87: Oort 1987, Wind85: Windhorst et al. 1985, Donn87: Donnelly et al. 1987.
\end{table}

\subsection{Ancillary radio data}

Deep radio observations of this field are available at 327, 610 and
1412 MHz using the WSRT and at 1400 and 4860 MHz with the VLA (See
Table 2). The 1412 MHz observations with the WSRT 3 km array have the
east-west beam of $\sim ~12.5^{''}$ and rms noise of 0.12 to 0.2
mJy/beam (Windhorst, Heerde \& Katgert, 1984) from a 12 hour
observation.  The resolution is comparable to that of GMRT at 150 MHz.
The second set of observations cover a smaller area of $\sim$ 0.7
degree$^2$, but reach  a deeper flux density limit of 0.3 mJy (Oort and
Windhorst, 1985).  A third set of deep observations at 1412 MHz of the
region has detected 349 sources down to 0.1 mJy (Oort, 1987). A total of 580
sources are found in the combined 1412 MHz catalogue. Deep observations
with the VLA at 1462 MHz were also carried out which yielded a little over
100 sources above a flux density limit of 0.23 mJy (Windhorst et
al. 1985).  The resolution of this VLA image was $\sim$ 15$^{''}$,
comparable to the WSRT resolution at 1412 MHz. The LBDS region was also mapped at
327 MHz using the WSRT with a resolution of about an arc-min.  Above a 5
$\sigma$ limit of 4.5 mJy, a total of 384 sources were found (Oort,
Seemers \& Windhorst, 1988).  The 610 MHz catalog (unpublished) has
$\sim$ 400 sources stronger than 1.3 mJy.  The VLA and WSRT
observations at 5 GHz were also used, though they cover a small
part of the area. Table 2 gives the summary of all available data, and
Figure 3 shows the areal overlap with the GMRT
observations. However, all these observations combined cover less than
50\% of the GMRT 150 MHz area. We also use the WENSS data at 327 MHz and NVSS
and FIRST data at 1400 MHz for the entire field.

\section{Results and discussion}

\subsection{The 150 MHz source catalog}

A catalogue of sources out to the 20\% of the peak primary beam response 
was created with the peak source brightness greater than 6
times the local rms noise value. The details of the source extraction
criterion are given in Sirothia et al. (2009). The final catalogue
contains about 750 sources above a flux density limit of $\sim$ 4 mJy. Since the
rms is not uniform across the image (rms is higher away from the phase
center and also near a few bright sources), the actual completeness
limit of this catalog is higher than this limiting value. The median flux
density of the catalogue is $\sim$ 26 mJy. The majority (about two-third)
of the sources are unresolved within the resolution of the
observation. About 18\% of the sources are significantly extended
(larger than about 0.5 arcmin). Close inspection of these sources
suggests a range of morphology, such as FR-II, FR-I, one sided jets, relic
types, etc.  Further details on these sources and their properties is
beyond the scope of this paper. The mean position offset at 150 MHz with
respect to the FIRST position is about -0.019$\pm$3.27 arc-sec in RA
and -1.61$\pm$3.12 arc-sec in DEC. A portion of the table is 
presented in Table 3 (the full catalogue is available online). 
Column 1: Source Name; Columns 2-4: RA; Columns 5-7: DEC; Columns
8-17: The flux density at 150 MHz(GMRT), 327 MHz(WSRT), 327 MHz(WENSS), 610 MHz(WSRT),
1412 MHz(WSRT, Windhorst et al. 1984), 1412 MHz(WSRT, Oort et al. 1987), 1400 MHz(NVSS), 1400 MHz(FIRST), 1462 MHz(VLA) and 4860 MHz(VLA) respectively;
Column 18: Spectral index from the fit.

The source count
reported from this field goes an order of magnitude deeper than earlier
150 MHz surveys (McGilchrist et al. 1990). In Fig. 4 we plot the GMRT
differential source count and for comparisons with previous surveys
we also plot the source count from 7C survey for the same region (Hales
et al. 2007 and references therein). We believe that our 150 MHz sources
are dominated by a cosmological population of AGNs.

\begin{figure}
\includegraphics[width=58mm,angle=270]{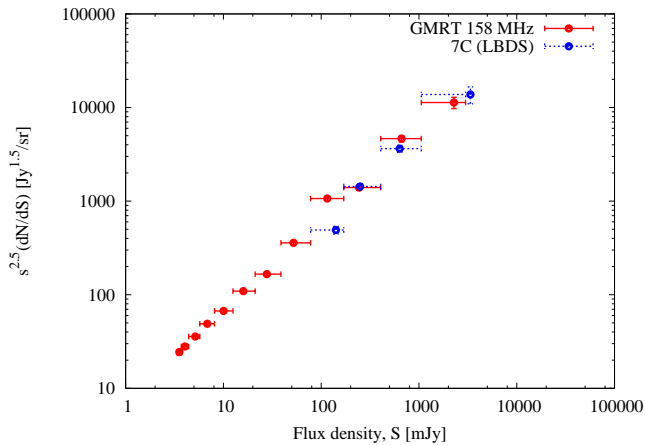}
\caption{LogN-LogS of the 150 MHz sources. We present sources extending
to an order of magnitude deeper than previous surveys. At higher
flux densities we are consistent with the
7C survey at this frequency (Hales et al. 2007). The source counts for 7C survey are shown for the
same region as that covered by the GMRT.}
\end{figure}

\begin{table*}
\caption{Sample table of GMRT 150 MHz sources (The complete table is available in online version).
The spectral index is from the
blind straightline fit, which may not  be valid for a small fraction ($\sim$ 3\%) of sources.}
\scriptsize
\begin{tabular}{@{\extracolsep{-5pt}}llllllllllllllllll}
NAME               & \multicolumn{3}{c}{RA (J2000)}  & \multicolumn{3}{c}{DEC (J2000)}   & S$_{150}$ & S$_{327}$   & S$_{WENSS}$      & S$_{610}$   & S$_{1412^a}$   & S$_{1412^b}$   &  S$_{NVSS}$     &  S$_{FIRST}$     & S$_{1462}$   &  S$_{4860}$   & $\alpha$\\
                   & hh & mm & ss.s & dd  & mm & ss.s & mJy     & mJy & mJy    &mJy  & mJy & mJy & mJy    &  mJy   & mJy & mJy  &      \\
(1)                & (2)& (3)& (4)  &(5)  &(6) &(7)   &  (8)    & (9) & (10)    & (11) &  (12)& (13) &  (14)   & (15)    & (16) & (17)  & (18)  \\
                   &    &    &      &     &    &      &         &     &        &     &     &     &        &        &     &      &      \\
GMRTJ083035+454329 & 08 & 30 & 35.7 &  45 & 43 & 29.6 & 11342.3 & $-$ & 3049.0 & $-$ & $-$ & $-$ &  614.4 &  608.3 & $-$ & $-$ & 1.48 \\
GMRTJ083752+445024 & 08 & 37 & 52.9 &  44 & 50 & 24.6 & 11158.2 & 4561.5 & 4662.0 & 2890.0 & 1519.2 & $-$ & 1528.9 & 1139.8 & $-$ & $-$ & 1.01 \\
GMRTJ084639+462141 & 08 & 46 & 39.7 &  46 & 21 & 41.9 &  3201.8 & 1299.4 & 1303.0 & $-$ & $-$ & $-$ &  362.3 &  380.0 & $-$ & $-$ & 1.03 \\
GMRTJ084737+461405 & 08 & 47 & 37.6 &  46 & 14 &  5.3 &  2819.6 & 1028.8 & 1043.0 & $-$ & $-$ & $-$ &  262.1 &  217.3 & $-$ & $-$ & 1.16 \\
GMRTJ084034+465112 & 08 & 40 & 34.3 &  46 & 51 & 12.8 &  2225.0 & $-$ &  928.0 & $-$ & $-$ & $-$ &  287.8 &  297.6 & $-$ & $-$ & 0.94 \\
GMRTJ084530+461035 & 08 & 45 & 30.6 &  46 & 10 & 35.5 &    47.1 &   28.0 &   19.0 & $-$ & $-$ & $-$ &    7.8 &    7.9 & $-$ & $-$ & 0.80 \\
\end{tabular}
\end{table*}

\subsection{Spectral index estimates}

The counterparts for the 150 MHz sources at higher radio frequencies were
searched within a 20${''}$ radius from the 150 MHz position. In addition to
the published deep observations of the LBDS field at 327, 610, 1412, 1462
and 4860 MHz, we used the WENSS catalog at 325 MHz and the NVSS and
FIRST catalogs at 1400 MHz to obtain the spectral index of sources found
at 150 MHz.  Since the aim of the project is to search for steep
spectrum sources, the spectral indices were computed using the GMRT 150
MHz catalog as the primary catalog. The spectral index was computed if
a counterpart was found at 610 MHz or at any of the higher frequencies. 
If the
counterpart was seen only at 327 MHz, the spectral index was not
computed, since the two frequencies are quite close by and the error
in the spectral index would be large.  Nonetheless, if the source was detected at
327 MHz either in WENSS or at the deep observation of this field at
327 MHz, this value of flux was also used to fit the spectrum
along with other available measurements.  The typical error on the
spectral index estimate was about 0.1, or better. A small fraction of
the sources (3\%) do not fit well with a straight spectrum and these are
mostly sources with spectral turnover.  Analysis of these sources is
beyond the scope of this paper.  The spectral index distribution is
presented in Fig. 6.  The full table is available online.

\begin{figure}
\includegraphics[width=59mm,angle=270]{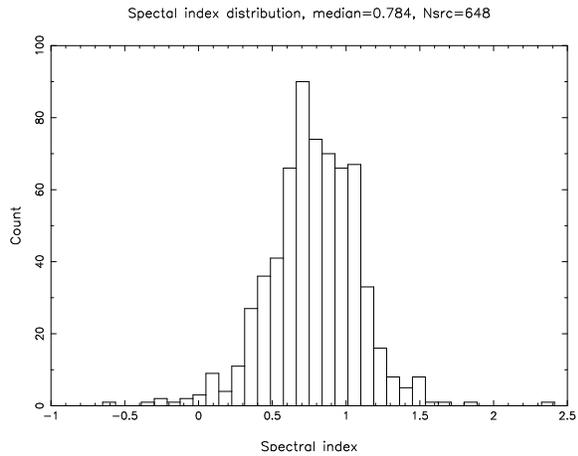}
\caption{Histogram of spectral index distribution of all sources with spectral index
measurements. The median value of spectral index is 0.78.}
\end{figure}

A total of 639 sources out of 765 (83\%) have spectral index determined.
The remaining 17\% sources are mostly
weak radio sources with a median flux density of $\sim$ 9 mJy, or fall
in the regions where deep observations at higher frequencies do not
exist.  The median spectral index of the sample is 0.78 (Fig. 5).
This is in good agreement with the similar measurements available in
the literature (e.g., Oort, Steemers and Windhorst, 1988; Gruppioni et
al. 1997). About 15 sources have spectra index flatter than 0.1, 100 sources
have spectra flatter than 0.5, 157 have a spectral index steeper than 1,
and about 20 have spectral index steeper than 1.3.  The steep spectrum sources are of
primary interest in this work, and we discuss these sources 
in the next section. Our sample of steep spectrum sources consists of
sources with spectra steeper than 1 (and not 1.3 as used by several
authors earlier) for the following reasons; (a) Despite the fact that 
most previous searches are limited to sources with spectral index
steeper than 1.3, the median spectral index of known HzRGs is 1.31.
(b) Though Klamer et al. (2006) have shown that the spectra for the
majority of HzRGs are straight over a large frequency range, evidence
for spectral curvature cannot be ruled out completely (Bornancini et
al. 2007). This means that at frequencies as low as 150 MHz, a higher
cut-off in the spectral index will translate into even higher cutoff
at the rest-frame of high-redshift objects.  (c) From the
$P-\alpha$ relation (Mangalam \& Gopal-Krishna, 1995; Blundell et al. 1999), sources which are less
powerful (at low radio frequencies) can have marginally flatter
spectra.  Since we are aiming to detect HzRGs that are 10 to 100 times
less luminous than the known HzRGs, we could miss HzRGs if we adopt a
steep spectral index cut-off of 1.3. Finally, our catalogue
will be cross-checked against SDSS for possible optical counterparts,
which would naturally eliminate any nearby objects that have steeper spectral index
for any other reason (e.g: synchrotron losses in galaxy clusters).
Therefore, we adopt a spectral index cutoff of 1, which yields a sample
of 157 radio sources selected at 150 MHz, which are then compared with
available deep optical observations.

\begin{figure}
\includegraphics[width=59mm,angle=270]{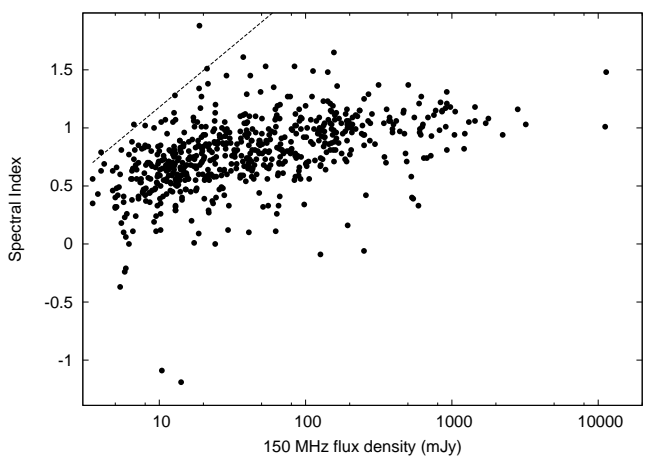}
\caption{The spectral index distribution of sources detected
at 150 MHz (S$_\nu \propto \nu^{-\alpha}$). The spectral index was computed whenever a 
counterpart was found at 610 MHz or higher frequencies, irrespective
of detection at 325 MHz. The dashed line shows the
spectral index limit corresponding to five sigma limit
of FIRST. There is a weak trend indicating that the fainter
sources tend to have flatter spectra.}
\end{figure}

\begin{figure*}
\vbox{
\hbox{
\includegraphics[width=40mm,angle=270]{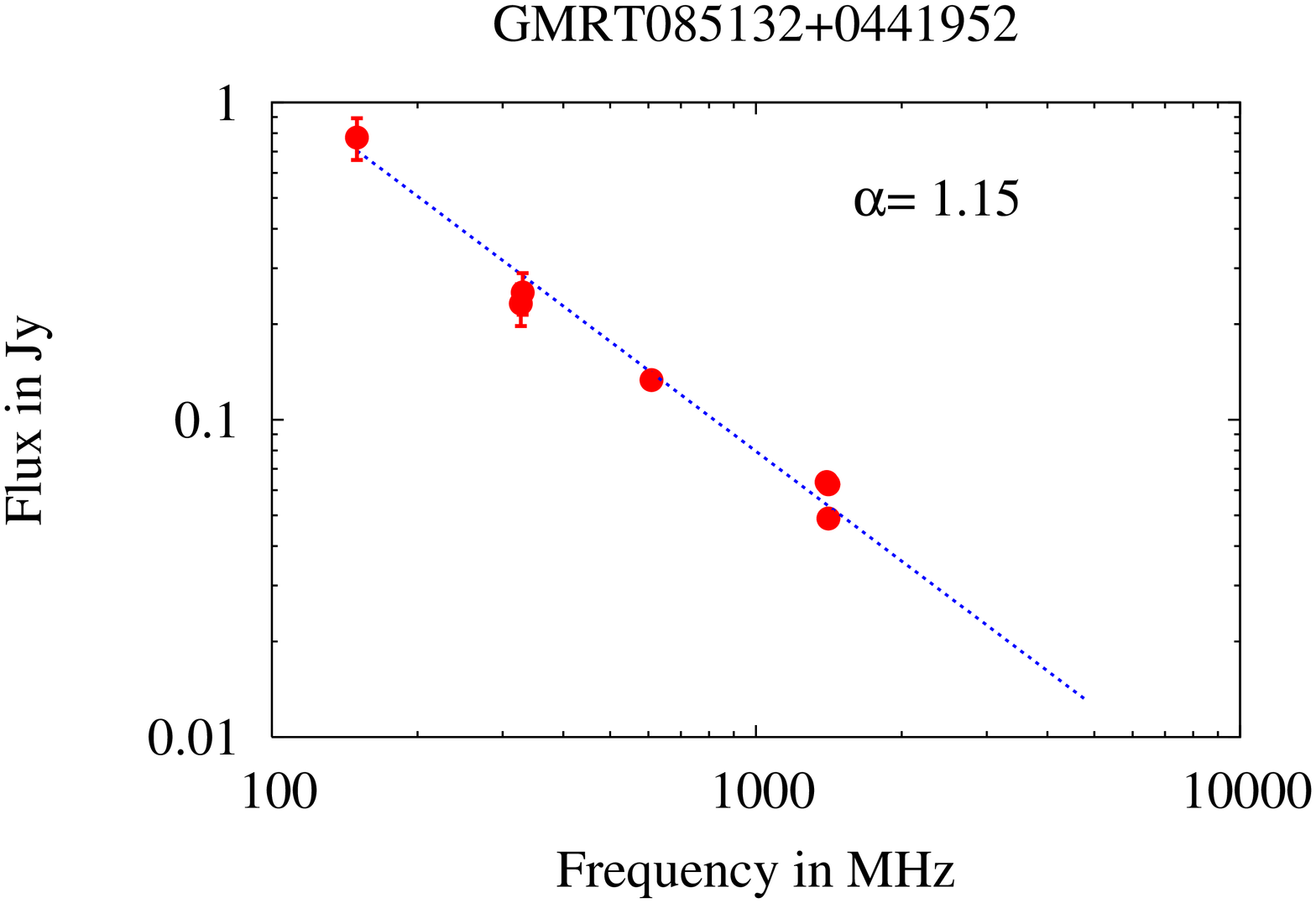}
\includegraphics[width=40mm,angle=270]{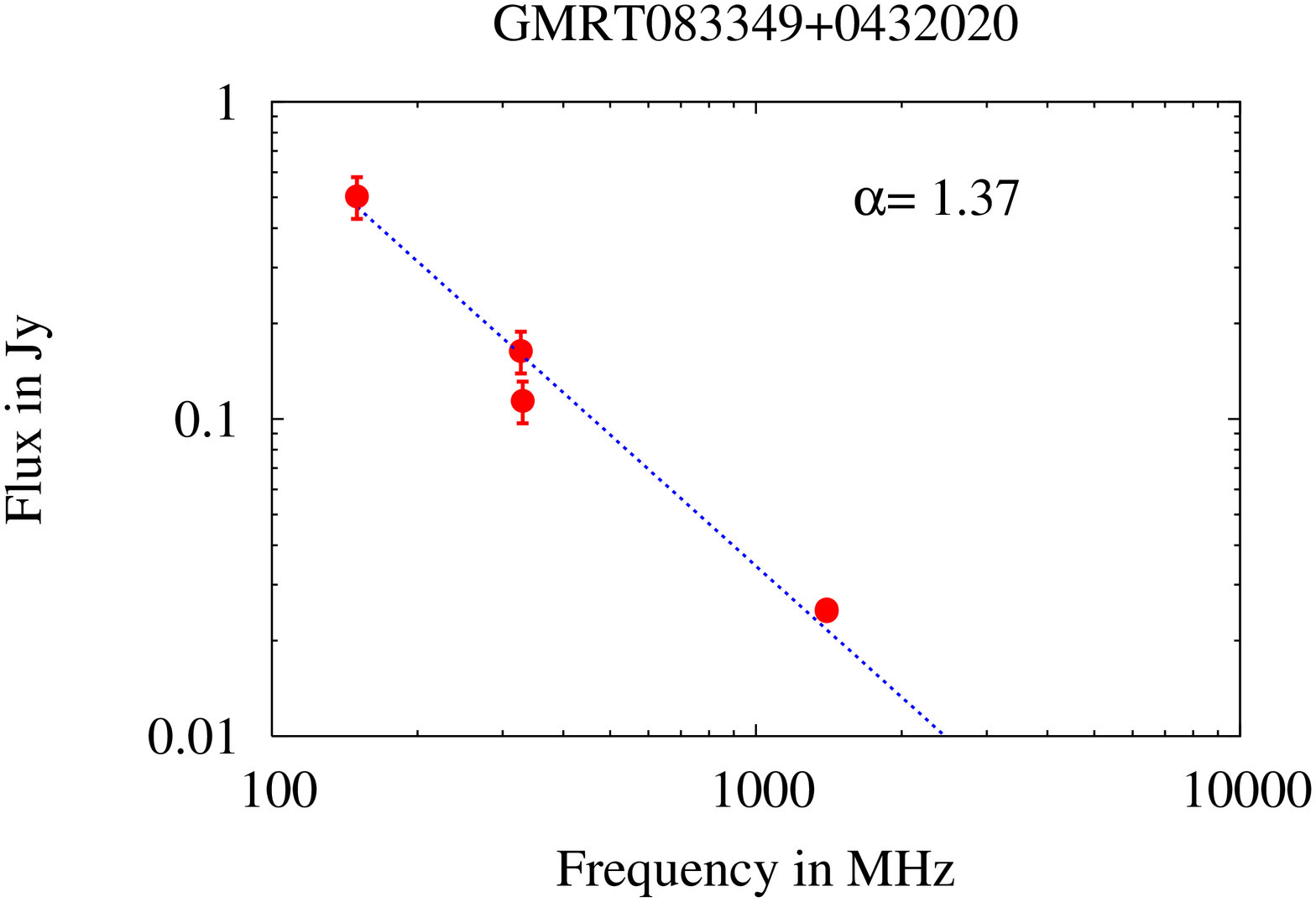}
\includegraphics[width=40mm,angle=270]{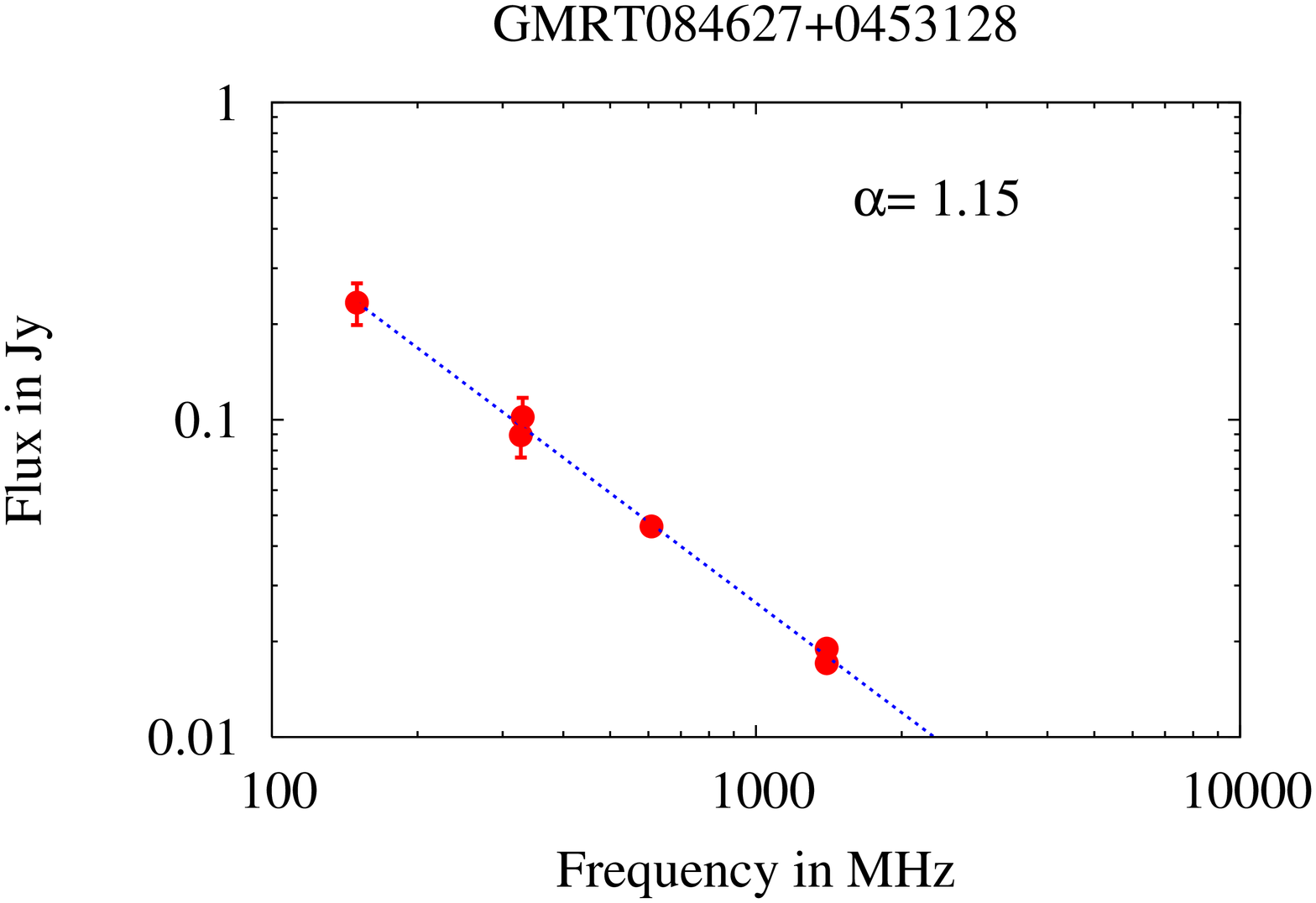}
}
\vspace{0.2in}
\hbox{
\includegraphics[width=40mm,angle=270]{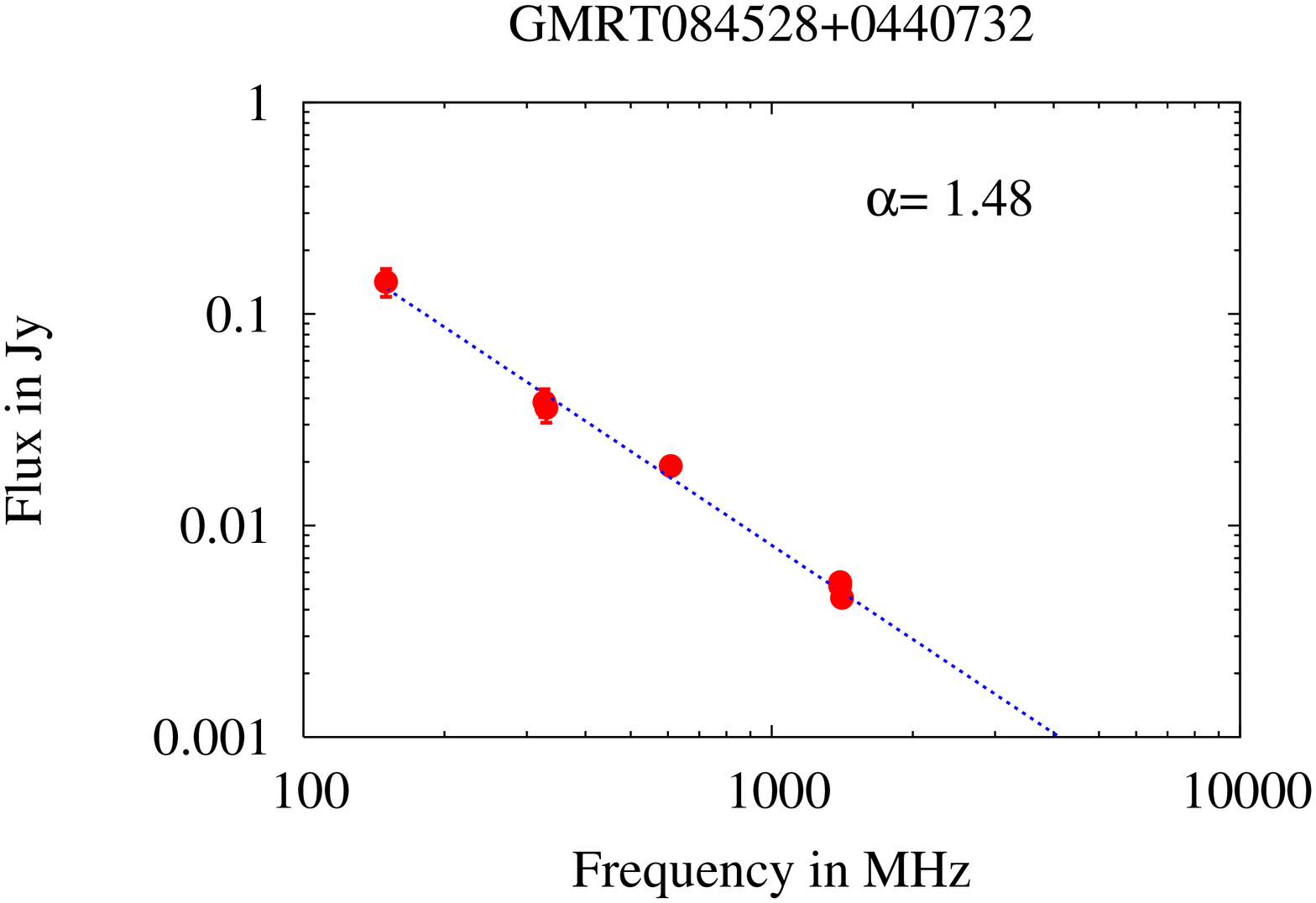}
\includegraphics[width=40mm,angle=270]{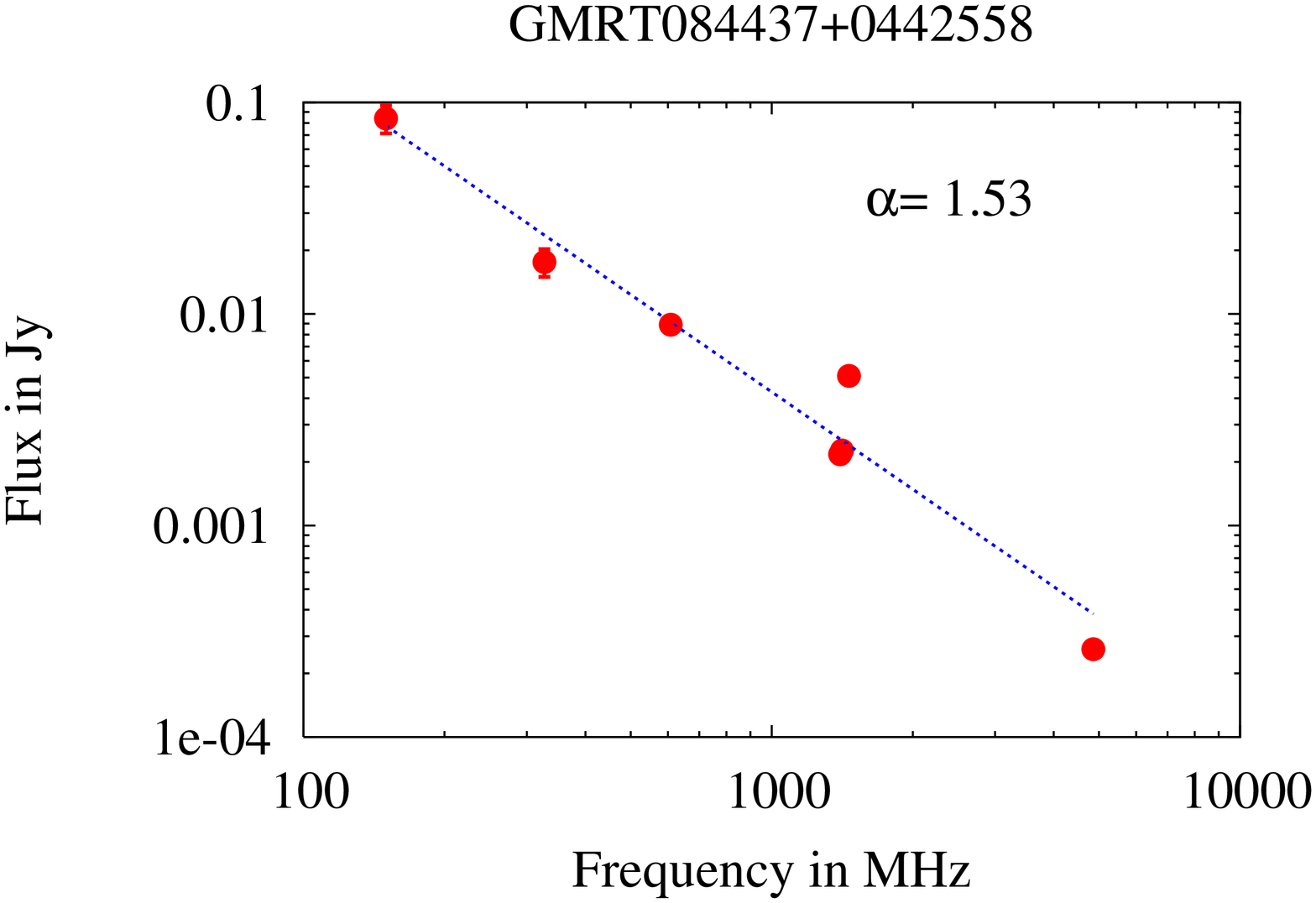}
\includegraphics[width=40mm,angle=270]{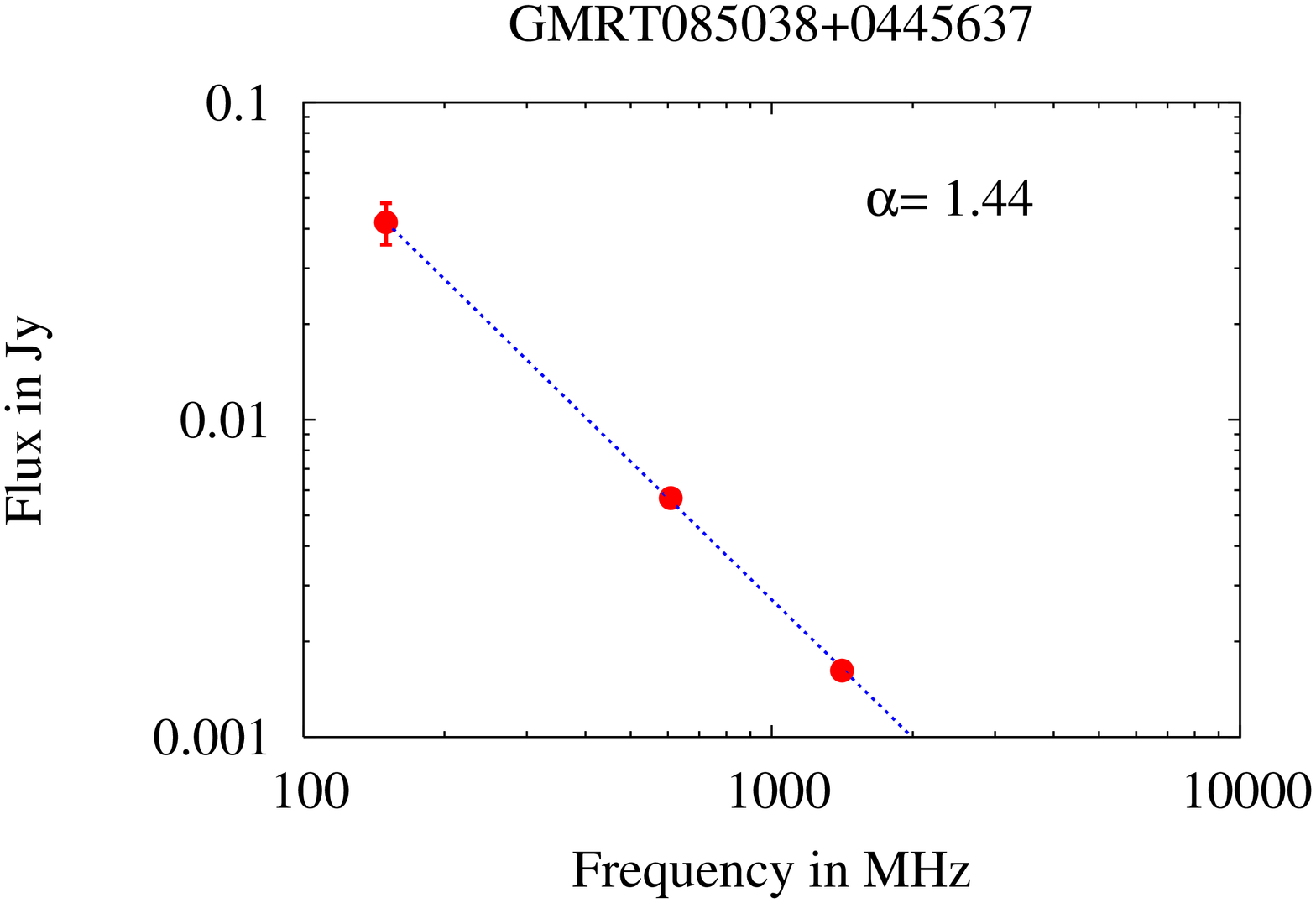}
}
\vspace{0.2in}
\hbox{
\includegraphics[width=40mm,angle=270]{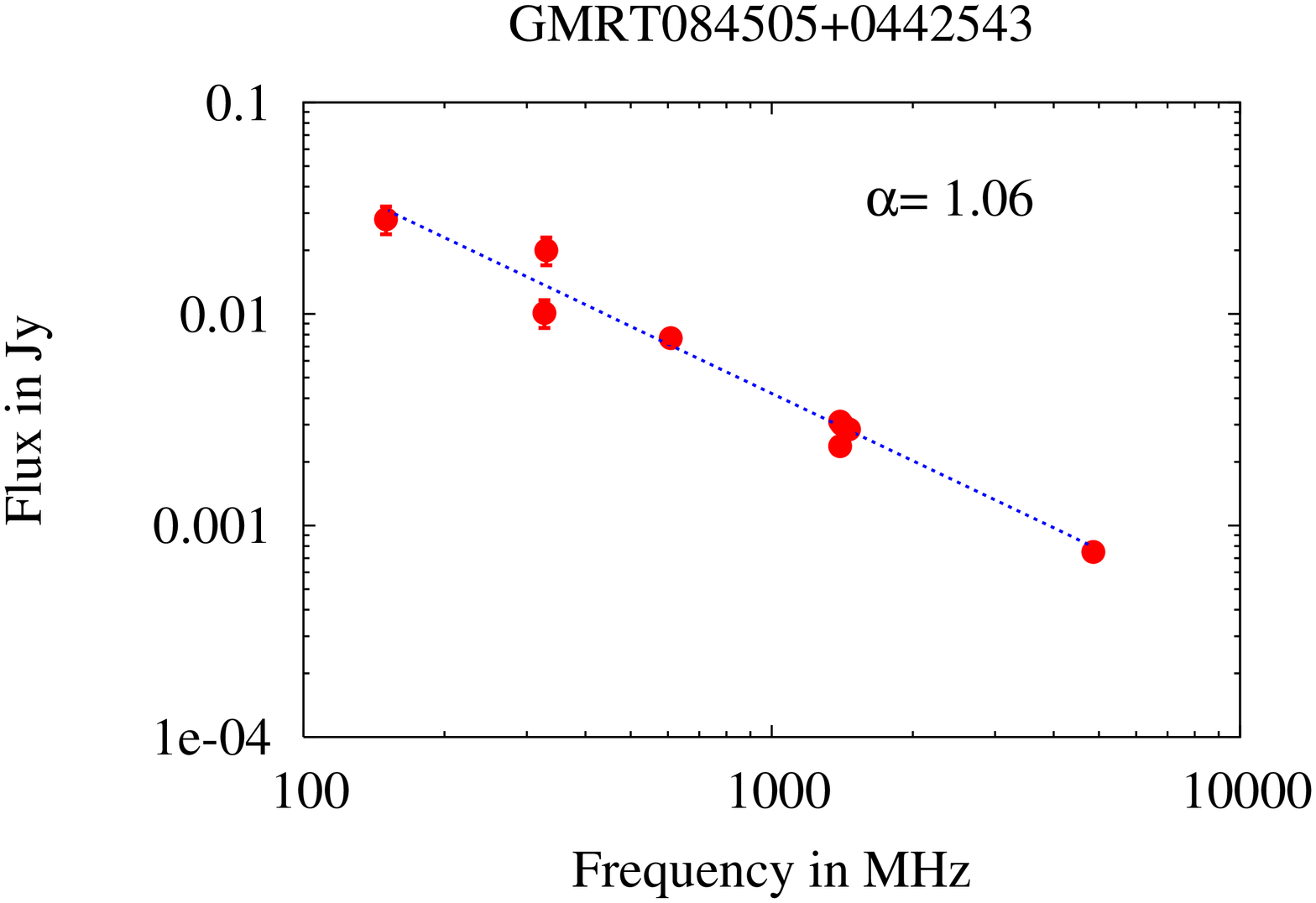}
\includegraphics[width=40mm,angle=270]{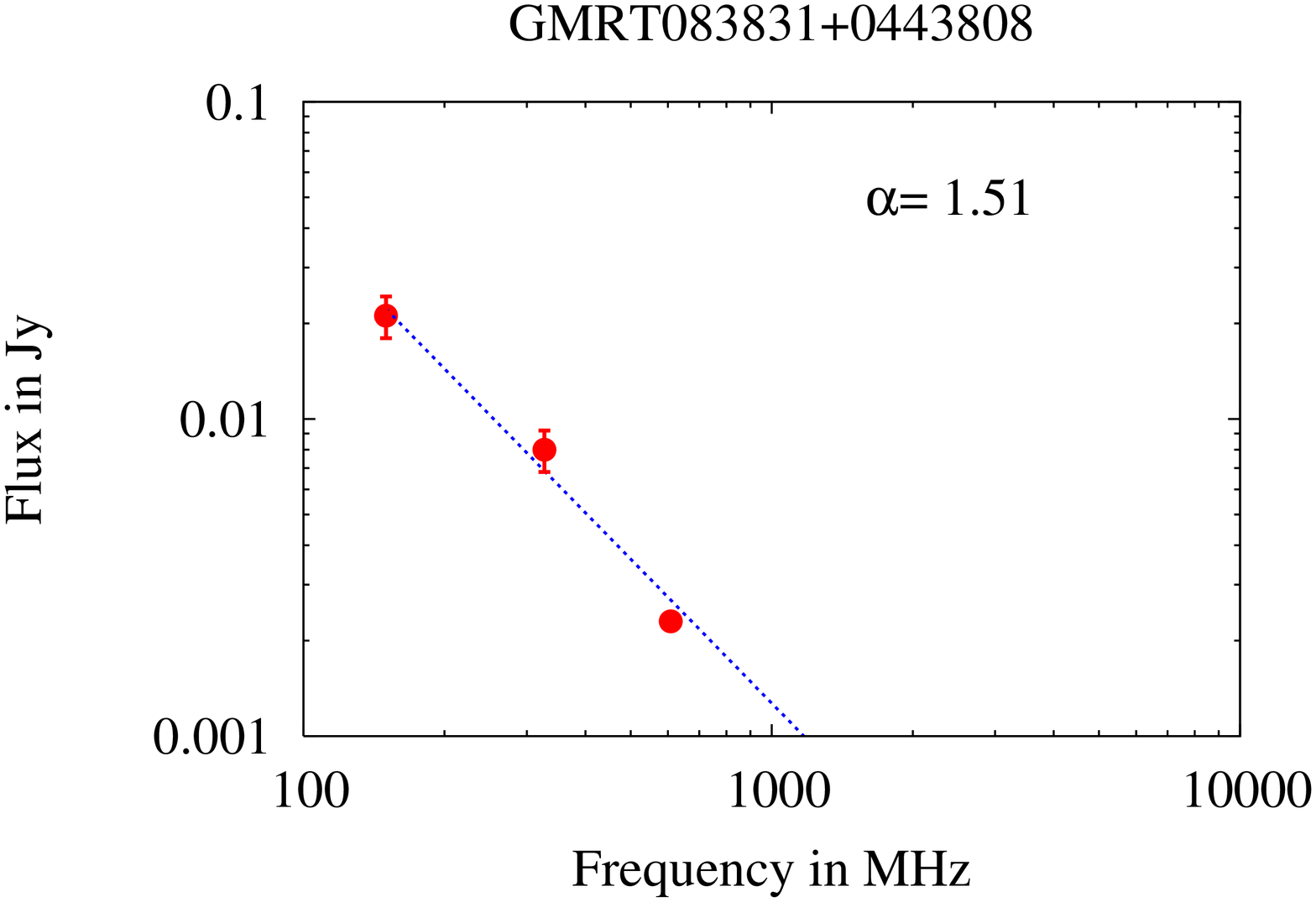}
\includegraphics[width=40mm,angle=270]{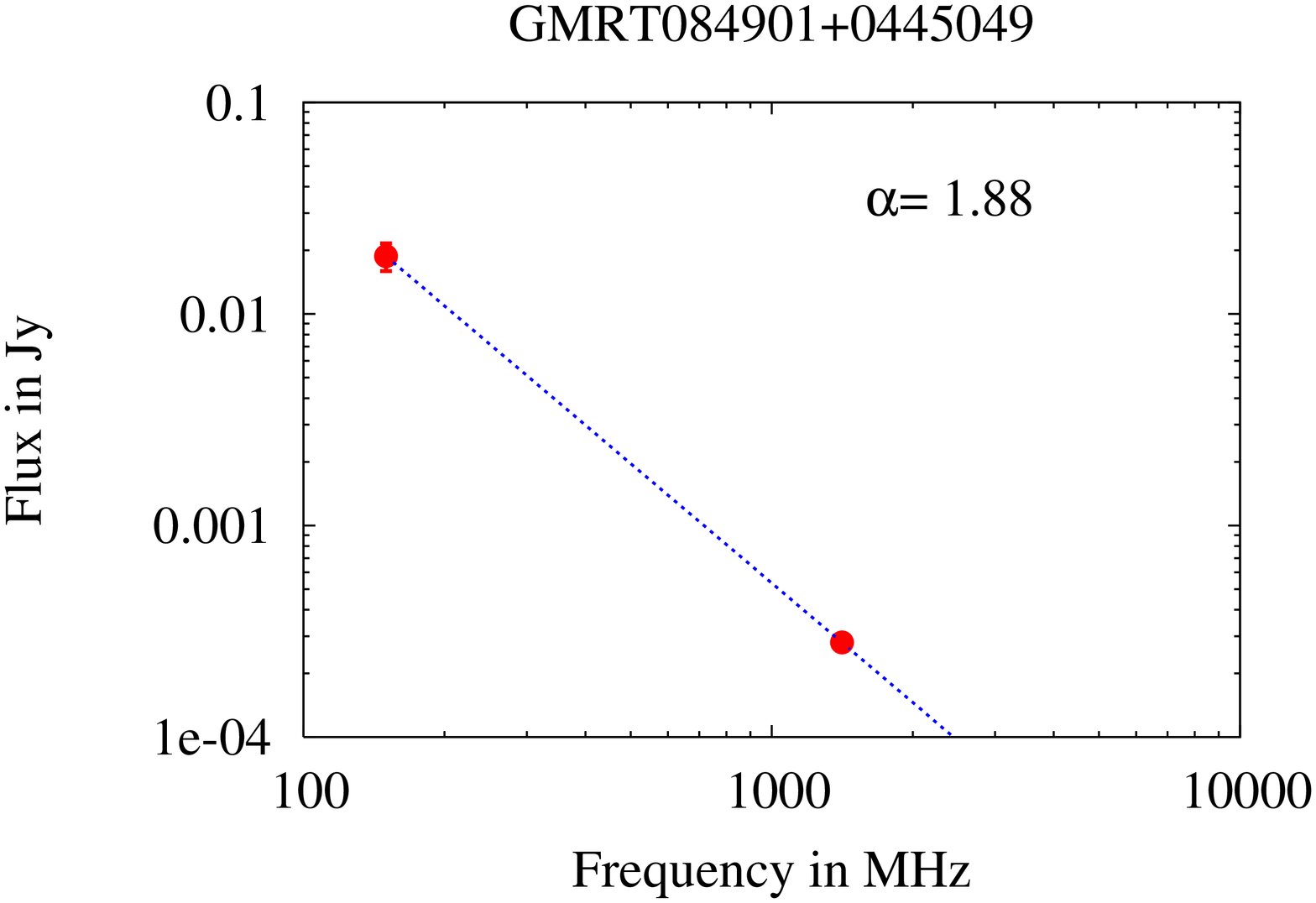}
}
}
\caption{Radio spectra of a few steep spectrum sources without an optical counterpart in SDSS. 
Note that for majority of the sources, the spectrum is straight over the frequency range 
from 150 to 1400 MHz.}
\end{figure*}

\subsection{Optical Counterparts in the Sloan Digital Sky Survey (SDSS)}

The position accuracy at 150 MHz is not good enough to search for
optical counterparts in SDSS. We have cross matched the sample of 157
sources, with spectral index steeper than 1 with VLA FIRST survey
(Becker et al 1995).  The position accuracy in FIRST survey is better
than an arcsecond.  If the counterpart is seen in FIRST, we have taken
the position from FIRST survey to obtain better accuracy. When multiple
components were seen in FIRST, the FIRST images were inspected and the
position of the core (or the centroid in case of a clear compact double
lobe structure) was chosen. In a couple of cases, the radio structures were such
that there was one unresolved compact component and a second component
which is extended. In such cases, we have taken the position of the
unresolved component. In this procedure, it is possible that we 
have missed likely counterparts in some cases. Among the 157 sources, 8
sources did not have counterparts in FIRST and hence the 150 MHz
position was used to cross match with SDSS.

We positionally matched each steep spectrum source position with the
photometric object catalog (PhotoObjAll) from the most recent (DR7)
release of the Sloan Digital Sky Survey (SDSS; Abazajian et al. 2009).
59 radio sources from this sample had at least one SDSS primary object
within 6 arcsec (38\% identification). The remaining 98 sources had no optical counterpart
in SDSS within 6${''}$ radius and these are listed in Table 4.
This radio source optical identification fraction is similar to that
found by Ivezic et al. (2002) (30\%) by cross-matching FIRST survey
sources with SDSS.

\subsubsection{Photometric redshifts}

The SDSS Catalog Archive Server (CAS) includes photometric redshift estimates
for most galaxies with flux measurements in multiple SDSS filters. Two
independent methods have been used to estimate the photometric redshift 1. A
template fitting method (Csabai et al. 2003) with outputs listed in
the table photoz of the CAS and 2. A neural network based method (Collister \&
Lahav 2004) with outputs listed in table photoz2 of the CAS. The neural network
based technique suffers from larger uncertainties and biases for
faint galaxies, because training examples with known spectroscopic
redshifts are not numerous enough for these objects. We, therefore,
chose to use only the photometric redshifts obtained by the template
fitting method. 44 of our 157 steep spectrum radio sources have a total
of 57 potential optical counterparts with listed photometric redhifts. Note that
the number of optical counterparts is greater than the number of steep
spectrum sources, because a few radio sources have more than one galaxy
with photometric redshift lying within the 6 arcsec search radius. The
photometric redshifts range from 0.02 to 0.75. The redshift histogram
for these galaxies is shown in Fig 8.

\begin{figure}
\includegraphics[width=85mm,angle=0]{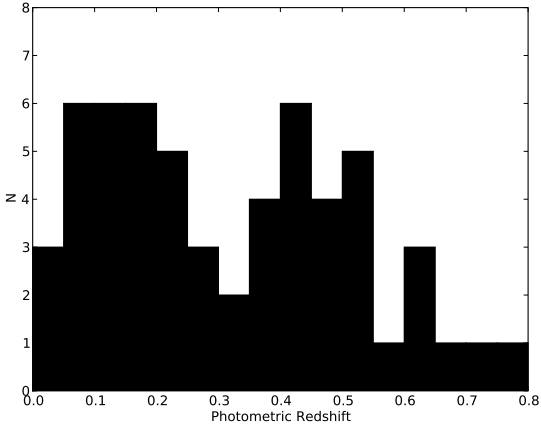}
\caption{The distribution of photometric redshifts of the steep spectrum
radio sources with optical counterparts in the SDSS.}
\end{figure}

\begin{figure*}
\vbox{
\hbox{
\includegraphics[width=45mm,angle=0]{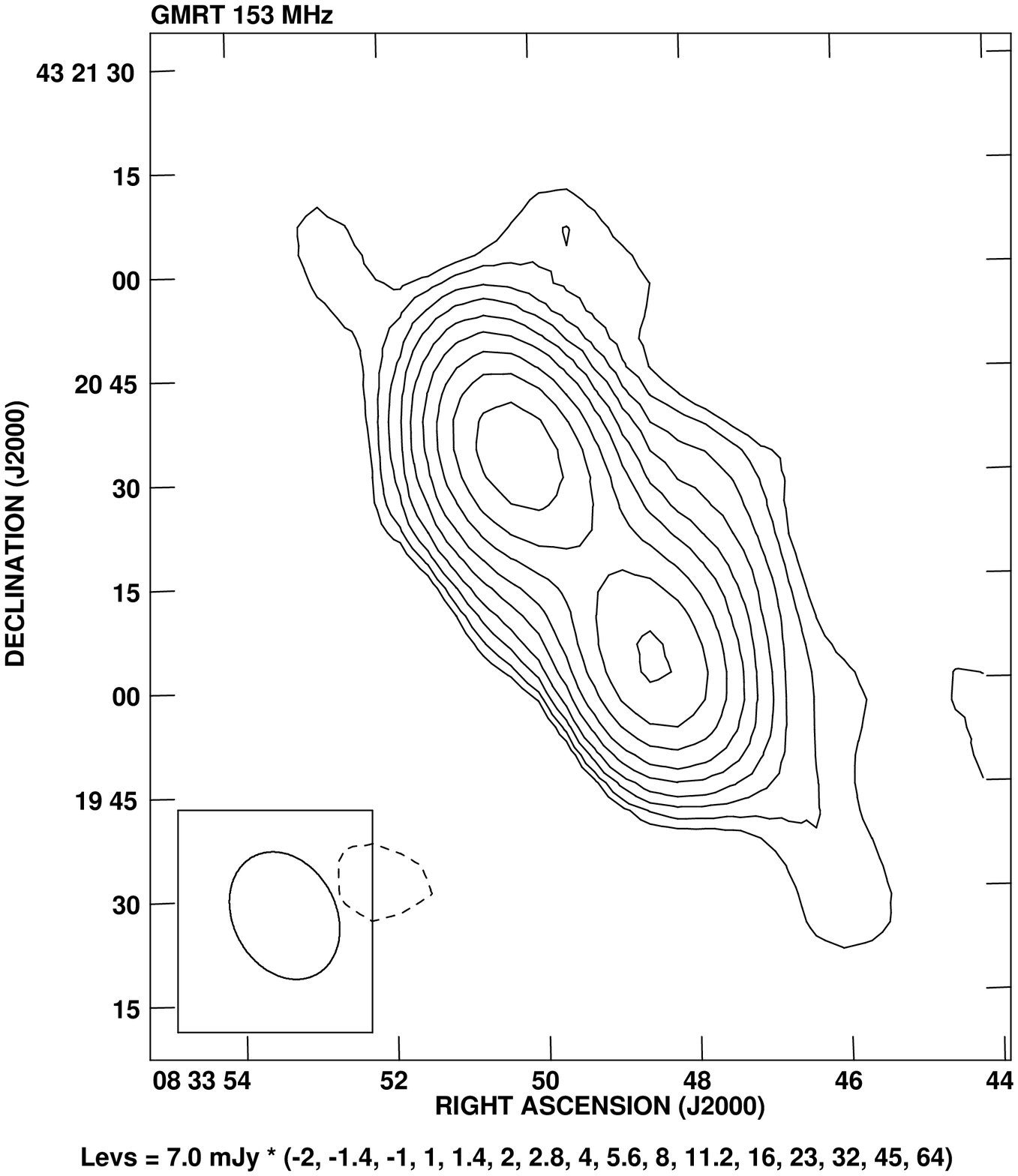}
\includegraphics[width=45mm,angle=0]{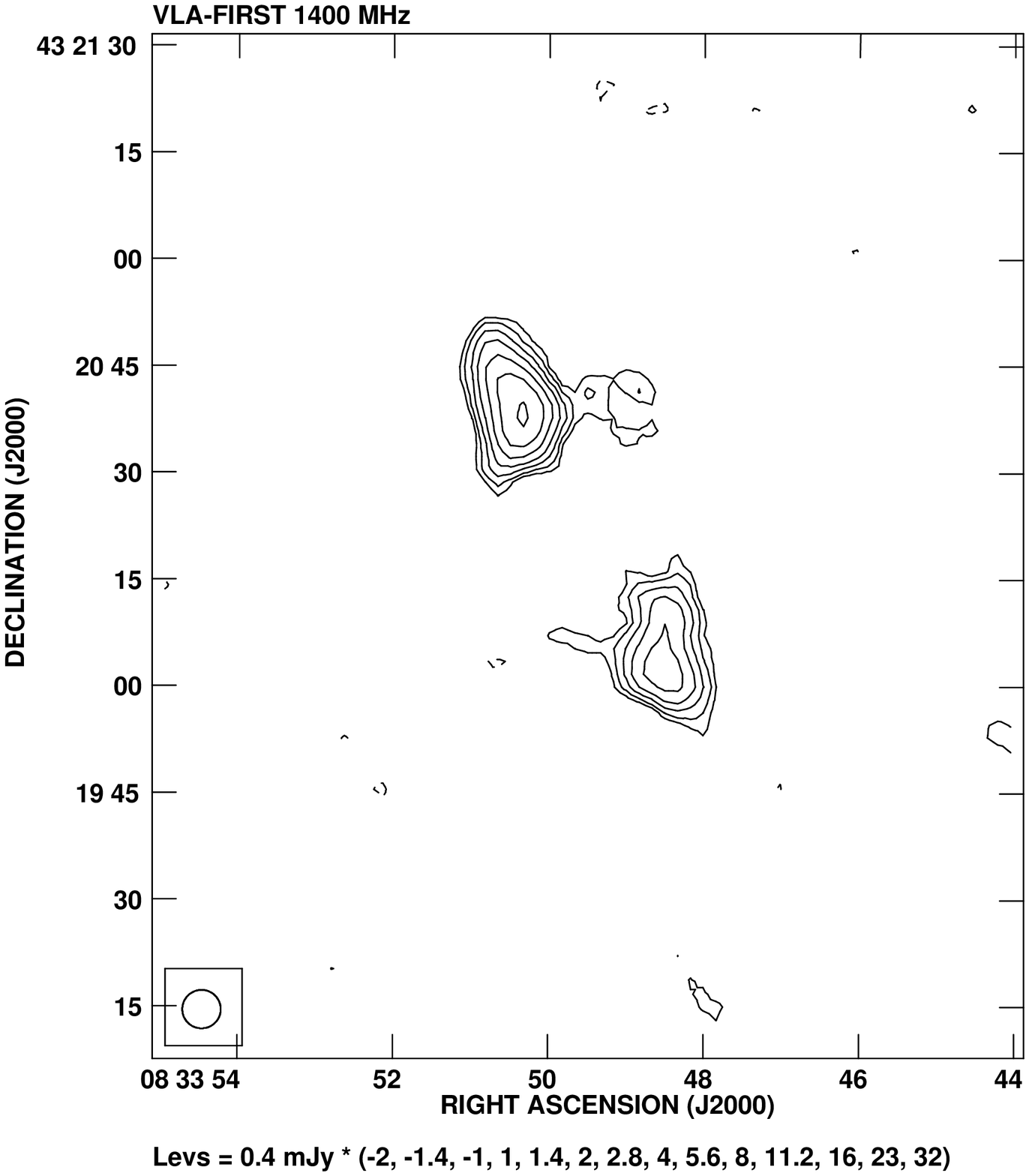}
\includegraphics[width=40mm,angle=0]{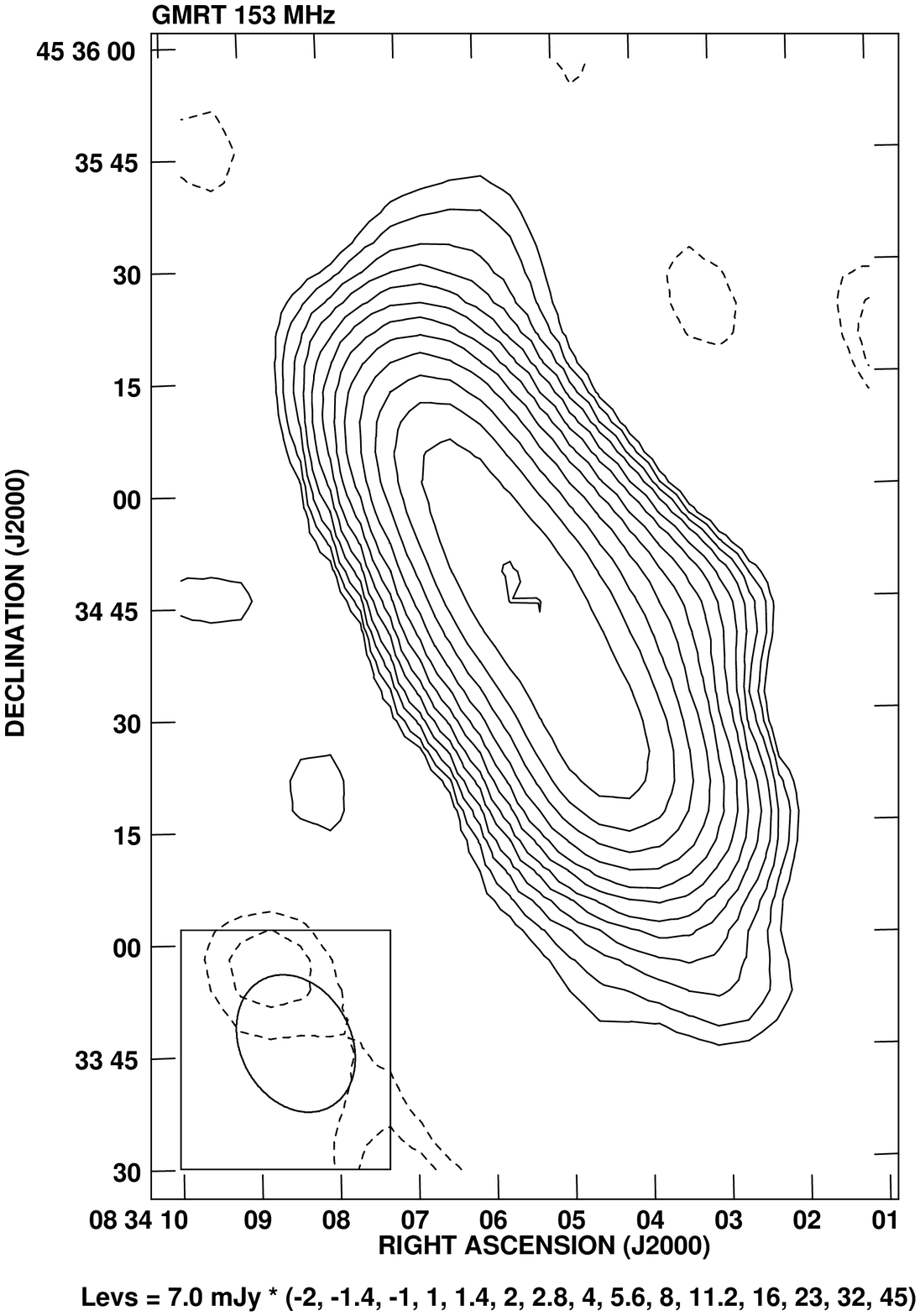}
\includegraphics[width=40mm,angle=0]{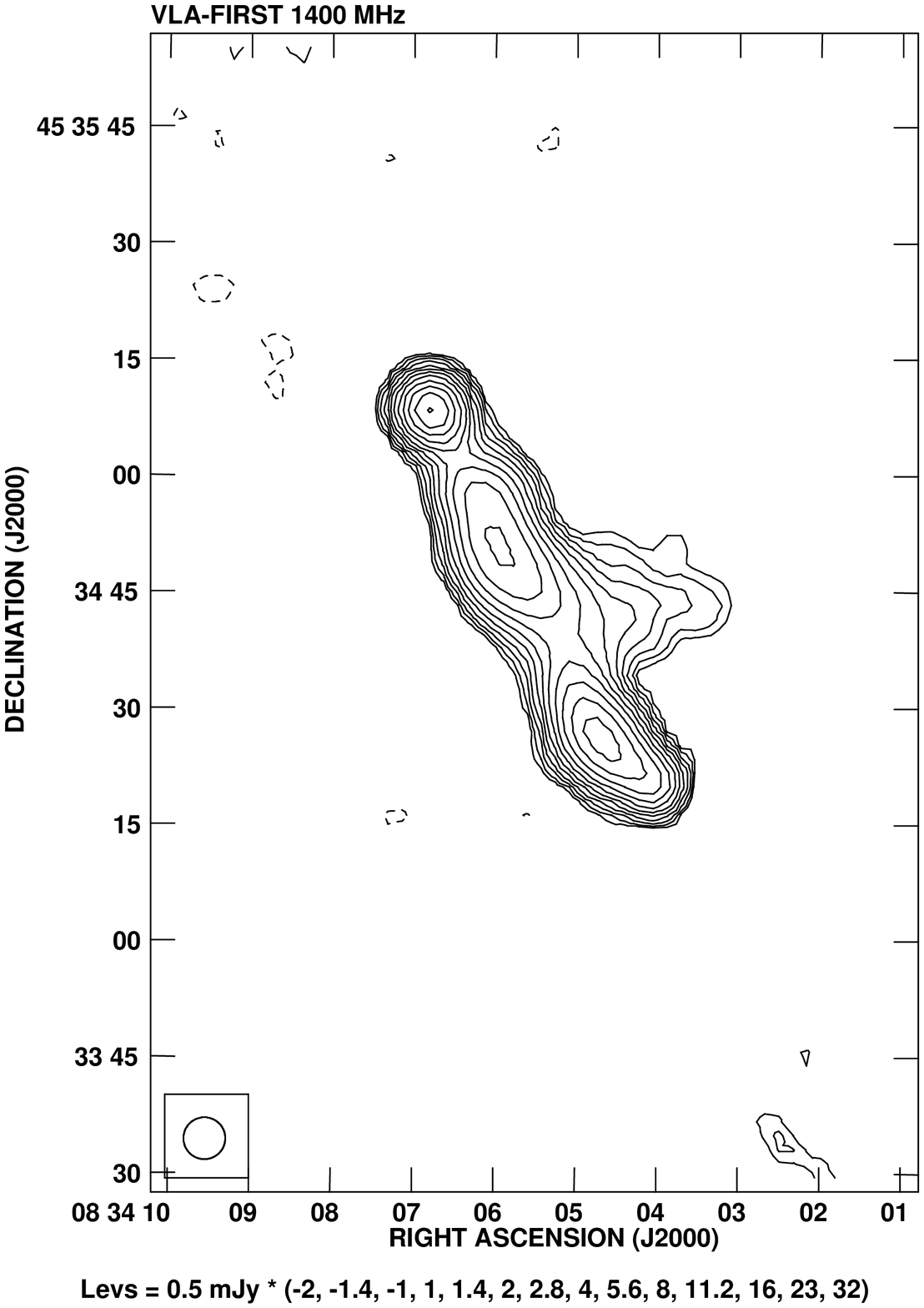}
}
\vspace{0.2in}
\hbox{
 \hbox{
\includegraphics[width=35mm,angle=0]{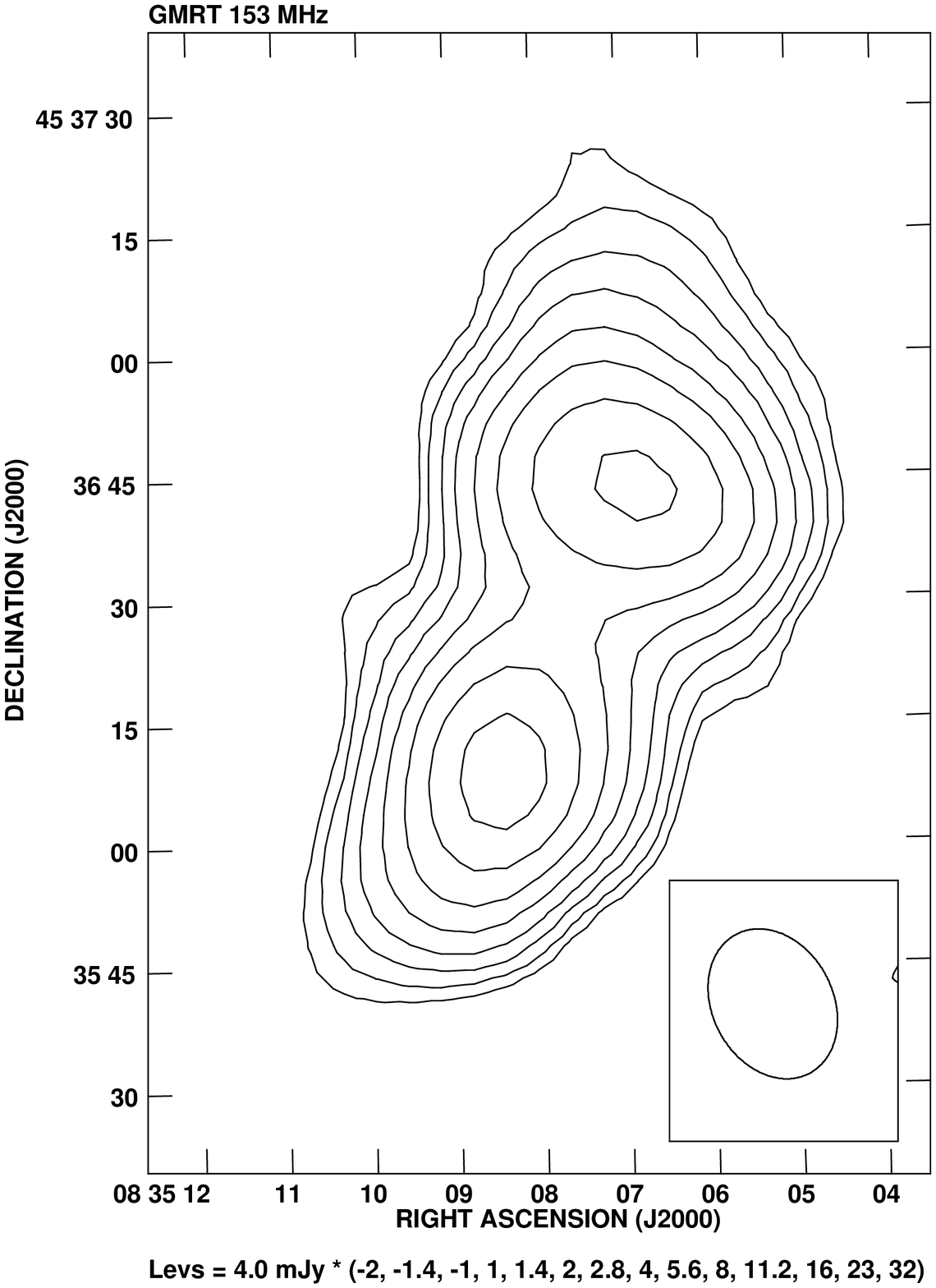}
\includegraphics[width=35mm,angle=0]{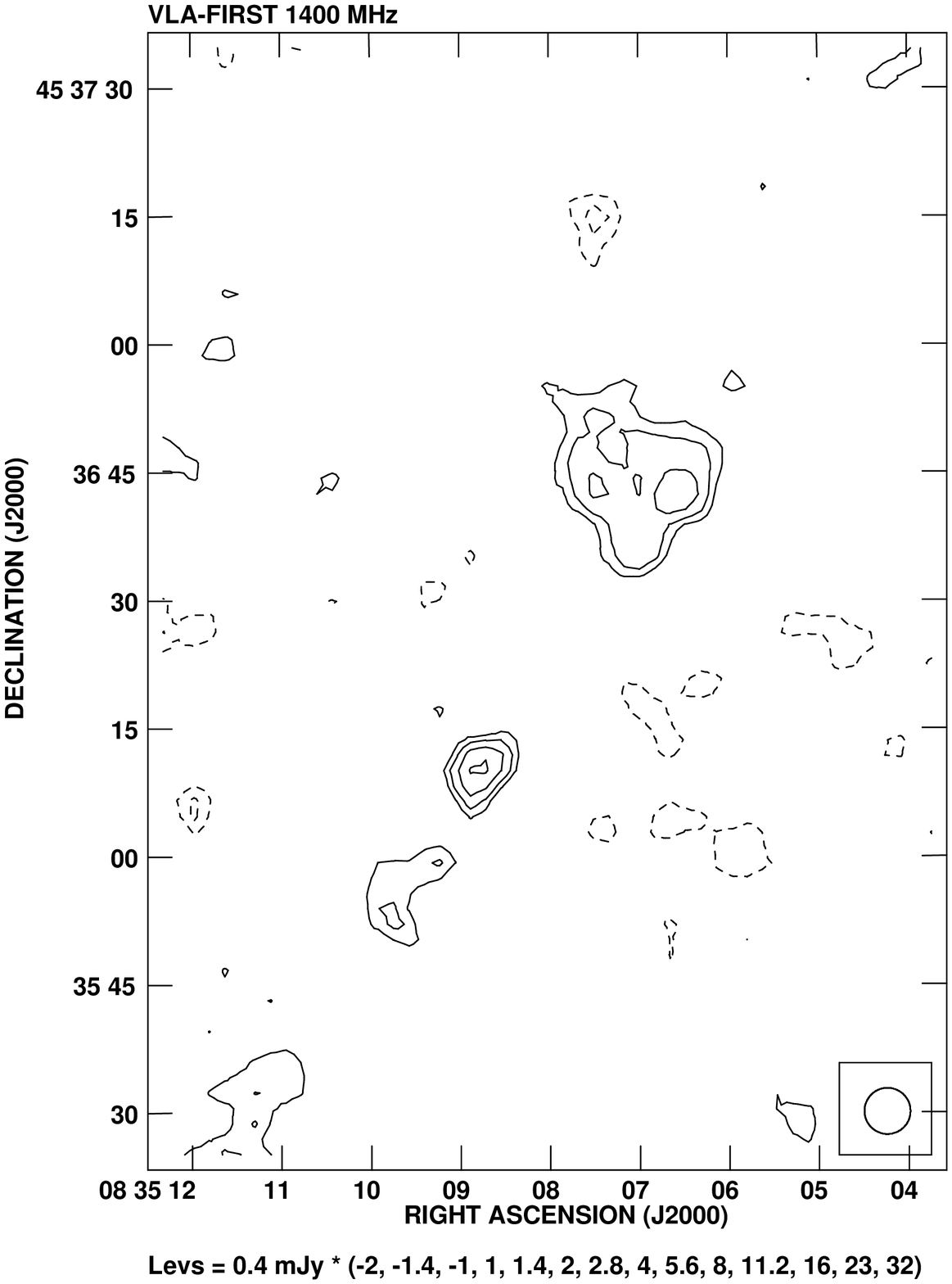}
\includegraphics[width=49mm, angle=0, viewport=-30 -40 650 600, clip]{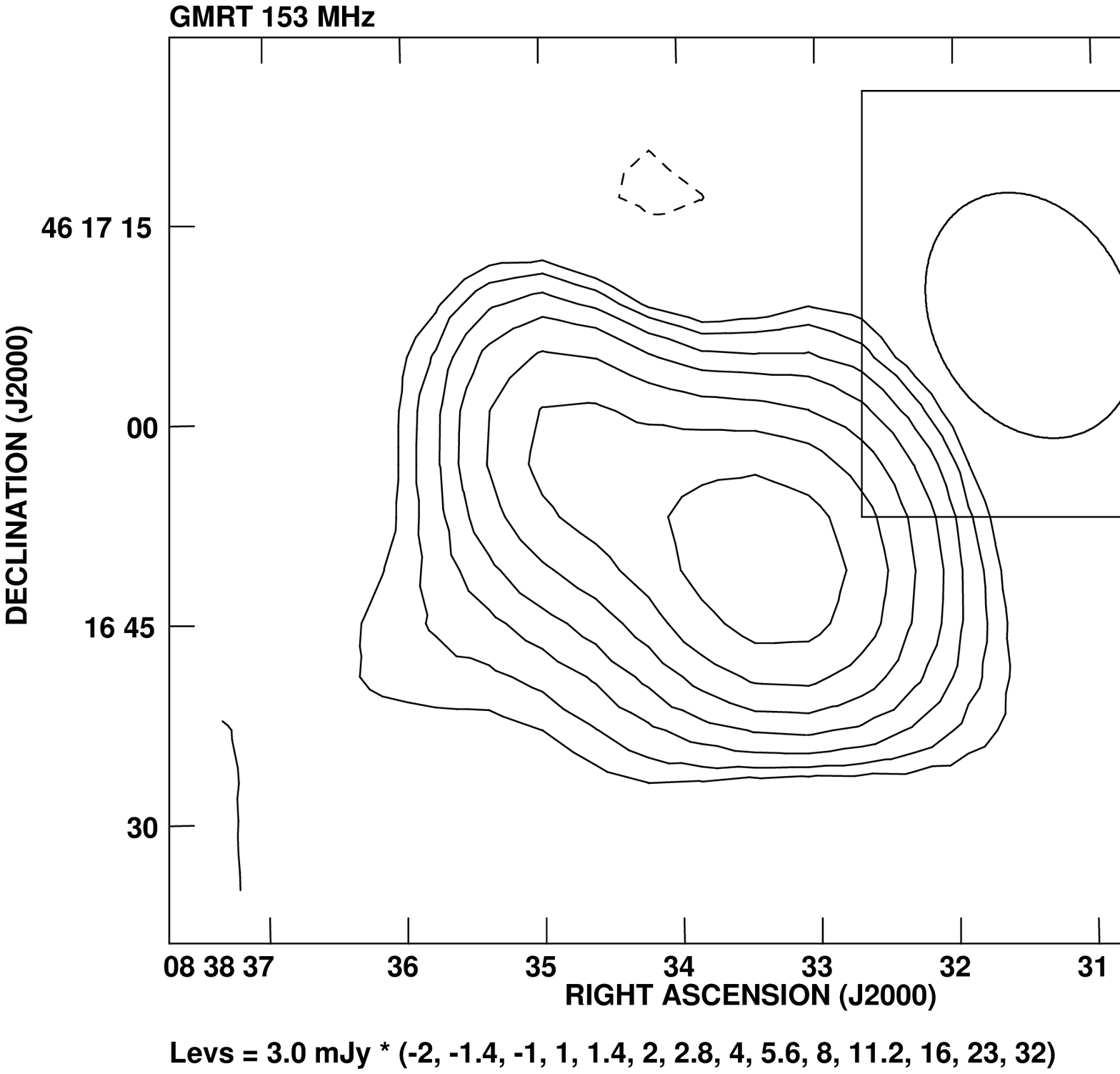}
\includegraphics[width=52mm, angle=0, viewport=-30 -10 700 600, clip]{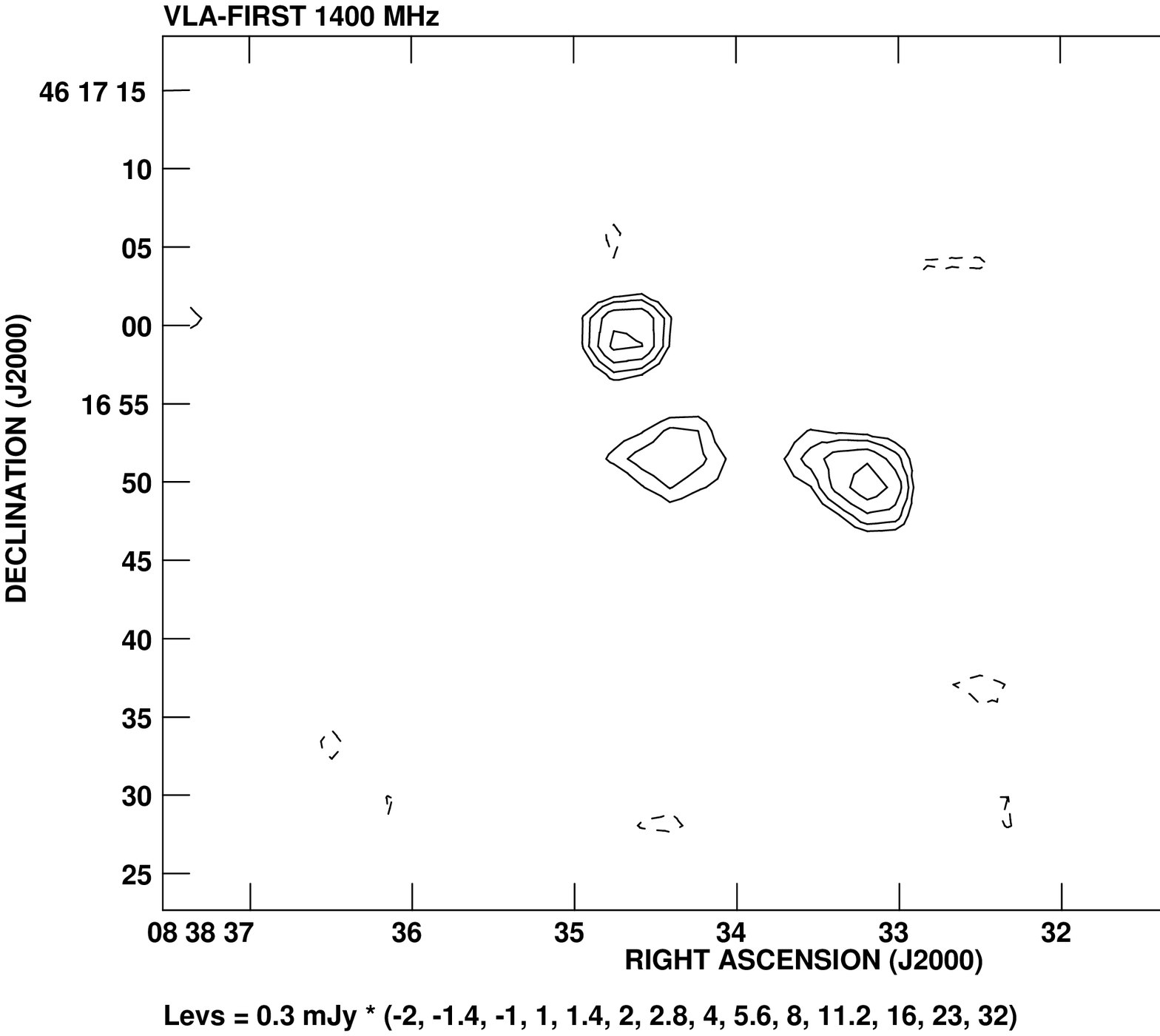}
  }
}
\vspace{0.1in}
\hbox{
\includegraphics[width=45mm,angle=270]{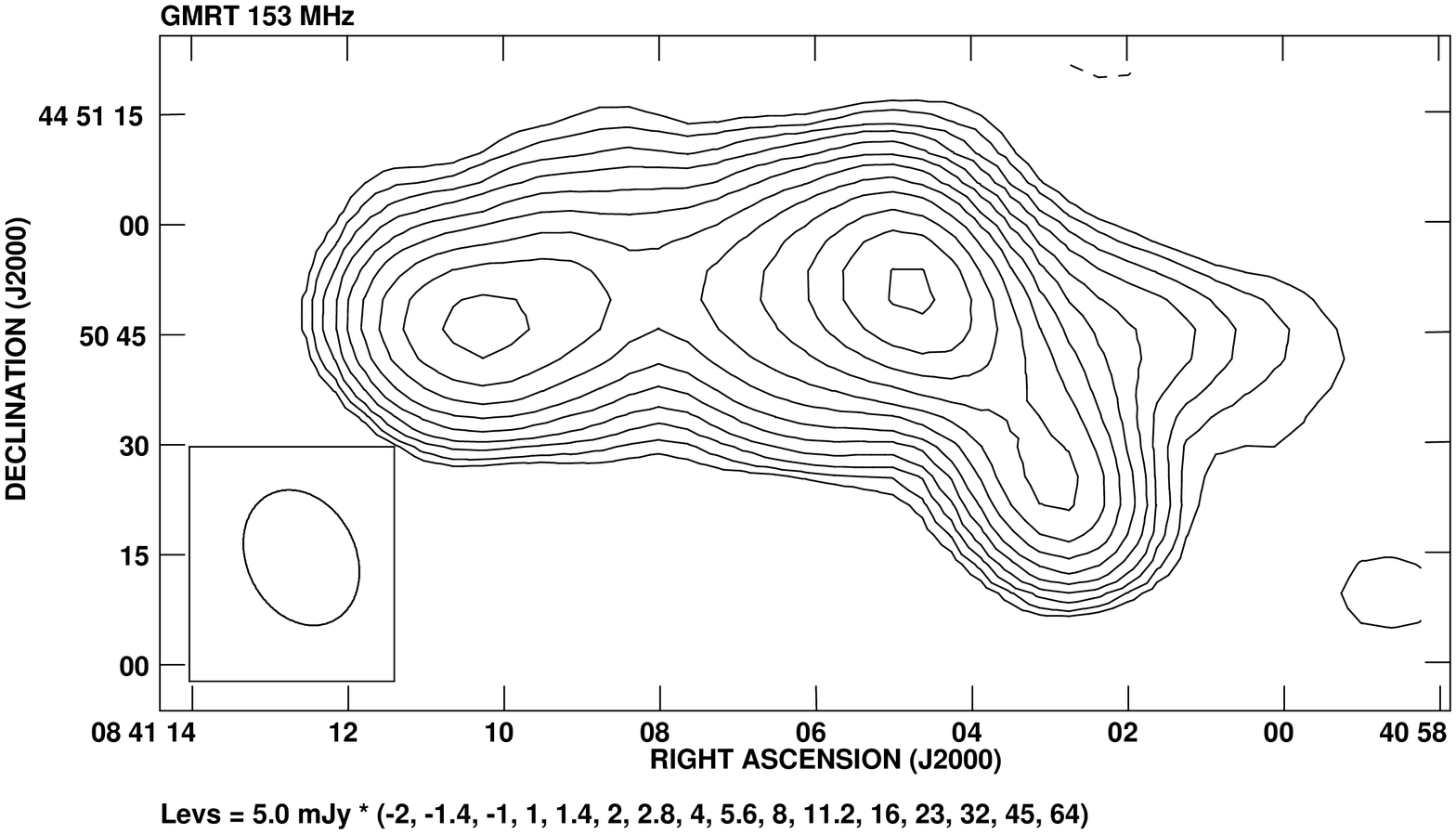}
\hspace{0.2in}
\includegraphics[width=45mm,angle=270]{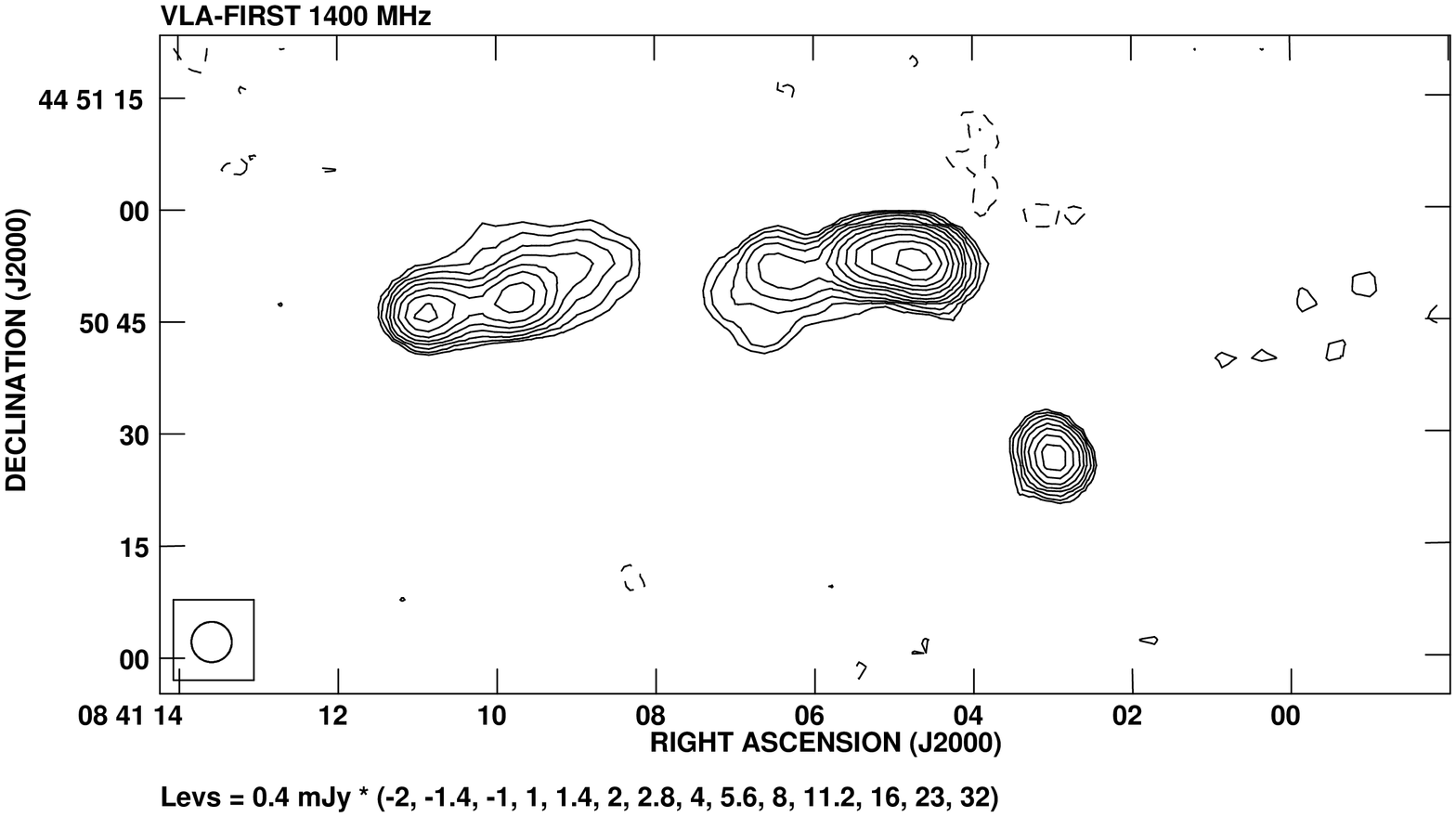}
}
\vspace{0.1in}
\hbox{
\includegraphics[width=45mm,angle=270]{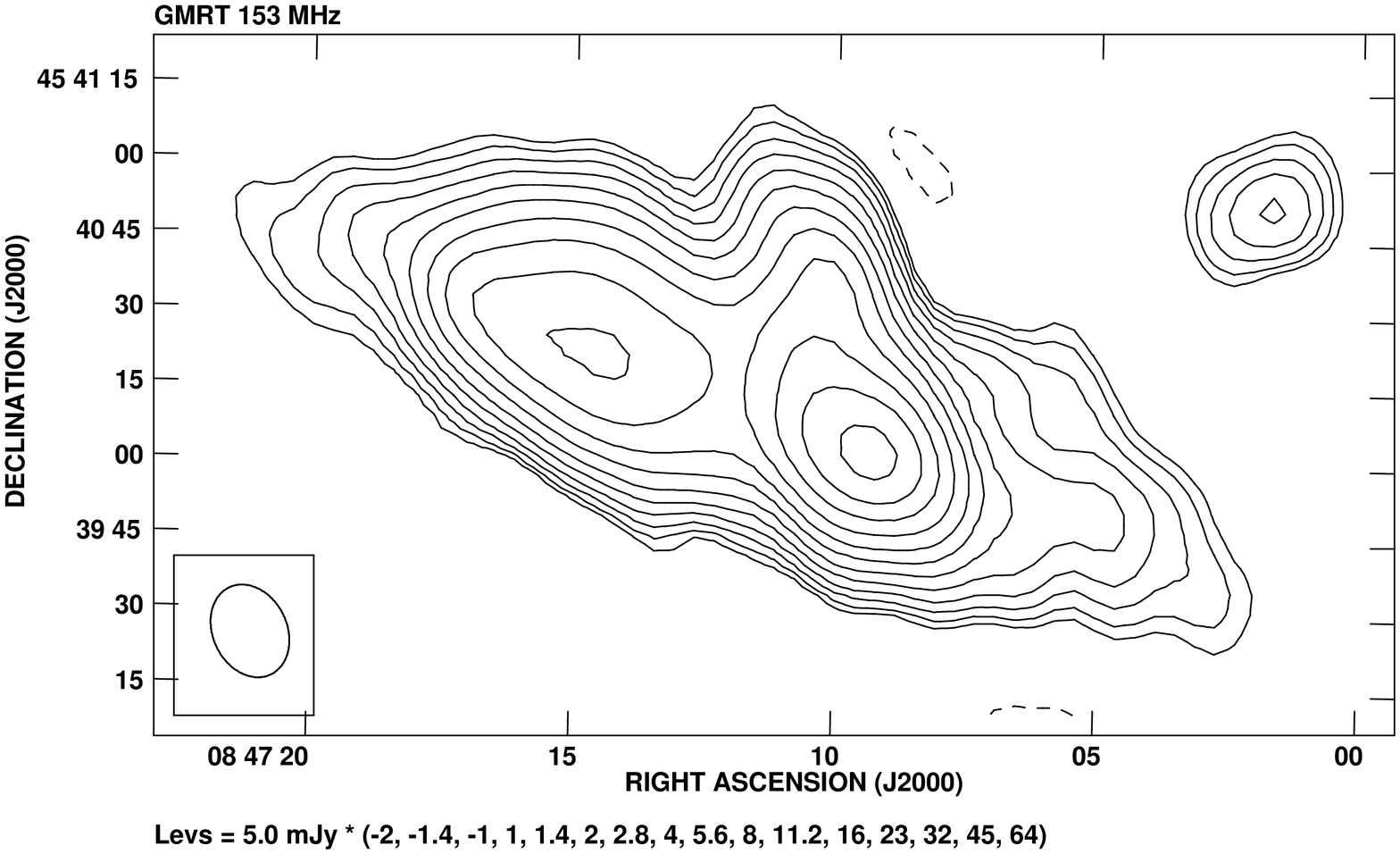}
\hspace{0.5in}
\includegraphics[width=45mm,angle=270]{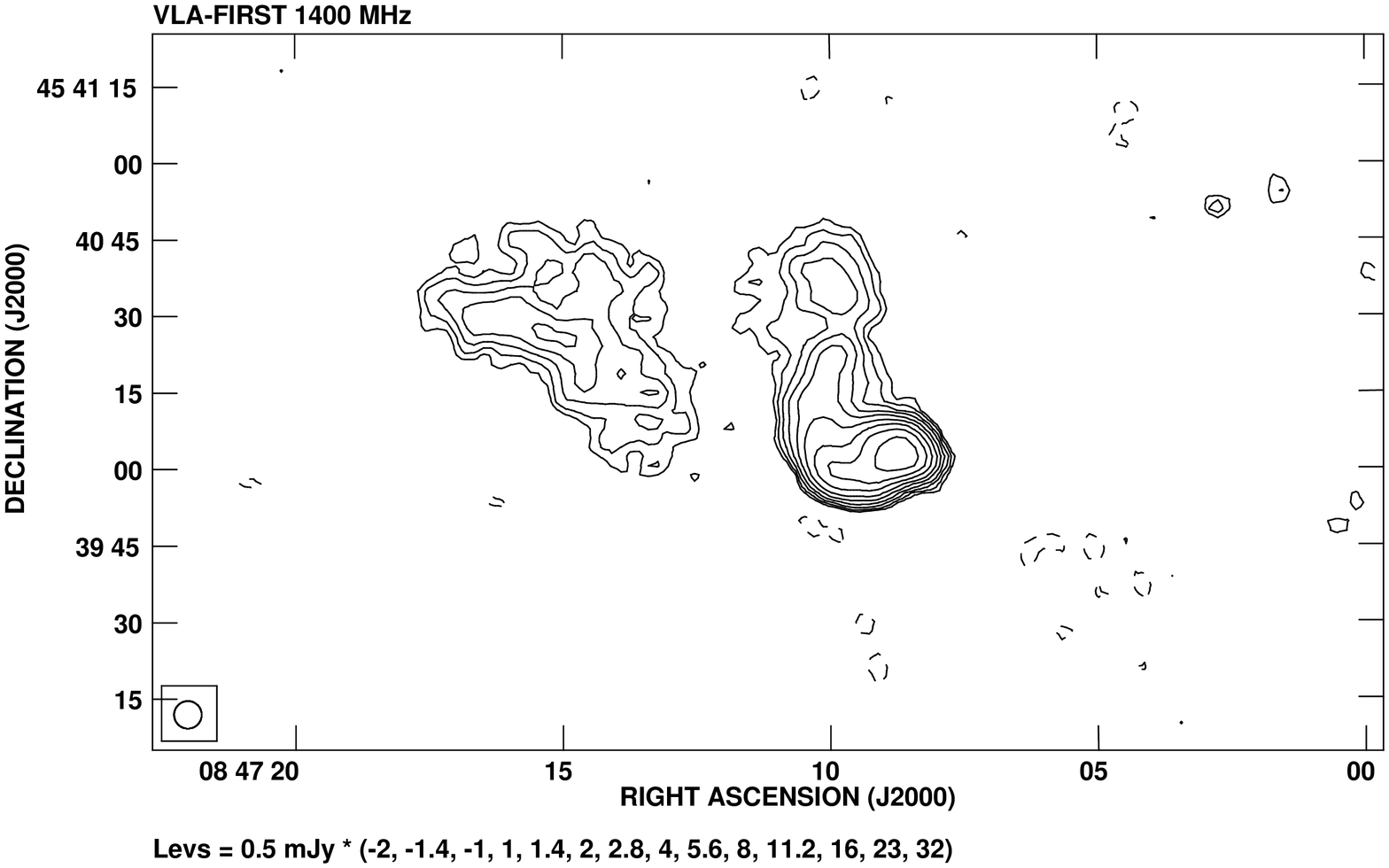}
}
}
\caption{A few extended sources among the steep spectrum source catalogue.
The GMRT 153 MHz and VLA 1400 MHz (FIRST) image pairs are arranged as follows:
Top line GMRTJ083349+432020 and GMRTJ083405+453443; second line
GMRTJ083507+453628 and GMRTJ083833+461651; third line GMRTJ084105+445045
and the image at the bottom is GMRTJ084711+454015 }
\end{figure*}

\begin{table}
\caption{Steep spectrum sources which do not have counterparts
in SDSS (see section 4.3).
The coordinates given are from the GMRT 150 MHz observations and the flux
density (S$_{150}$) is in mJy.}
\begin{tabular}{lllrl}
                       &               &                &          &        \\
NAME                   &  RA           &  DEC           & S$_{150}$ & $\alpha$ \\
                       &               &                &          &        \\
GMRTJ082947+442036 & 08 29 47.7  &  44 20 36.5  &   975.8  &   1.18 \\
GMRTJ083132+450742 & 08 31 32.4  &  45 07 42.8  &    55.9  &   0.99 \\
GMRTJ083156+435728 & 08 31 56.3  &  43 57 28.1  &  1054.0  &   1.14 \\
GMRTJ083212+451934 & 08 32 12.3  &  45 19 34.2  &    90.8  &   1.06 \\
GMRTJ083245+450716 & 08 32 46.0  &  45 07 16.2  &    75.5  &   1.27 \\
GMRTJ083349+432020 & 08 33 49.3  &  43 20 20.4  &   503.1  &   1.37 \\
GMRTJ083405+453443 & 08 34 05.3  &  45 34 43.0  &  1776.4  &   1.08 \\
GMRTJ083421+441057 & 08 34 21.2  &  44 10 57.7  &   905.8  &   1.17 \\
GMRTJ083507+453628 & 08 35 07.7  &  45 36 28.6  &   251.2  &   1.25 \\
GMRTJ083521+453305 & 08 35 22.0  &  45 33 05.3  &   416.8  &   1.06 \\
GMRTJ083527+440950 & 08 35 27.3  &  44 09 50.3  &   596.5  &   1.21 \\
GMRTJ083527+440004 & 08 35 27.4  &  44 00 04.0  &    94.2  &   1.08 \\
GMRTJ083548+450944 & 08 35 48.8  &  45 09 44.5  &    12.4  &   1.05 \\
GMRTJ083620+444826 & 08 36 20.7  &  44 48 26.3  &    17.2  &   1.06 \\
GMRTJ083628+430053 & 08 36 29.0  &  43 00 53.0  &    64.2  &   1.19 \\
GMRTJ083632+450605 & 08 36 32.8  &  45 06 05.4  &    41.3  &   1.10 \\
GMRTJ083701+435554 & 08 37 01.9  &  43 55 54.5  &    12.6  &   1.13 \\
GMRTJ083714+462908 & 08 37 14.8  &  46 29 08.9  &    18.2  &   1.05 \\
GMRTJ083757+443443 & 08 37 57.0  &  44 34 43.4  &   378.6  &   1.06 \\
GMRTJ083808+441910 & 08 38 08.5  &  44 19 10.1  &    39.6  &   1.29 \\
GMRTJ083814+424248 & 08 38 14.0  &  42 42 48.2  &    20.8  &   1.01 \\
GMRTJ083816+460632 & 08 38 16.5  &  46 06 32.7  &    39.3  &   1.04 \\
GMRTJ083825+432522 & 08 38 25.6  &  43 25 22.2  &   315.1  &   1.37 \\
GMRTJ083831+443808 & 08 38 31.5  &  44 38 08.2  &    21.2  &   1.51 \\
GMRTJ083833+461651 & 08 38 33.8  &  46 16 51.6  &    53.3  &   1.53 \\
GMRTJ083840+460825 & 08 38 40.5  &  46 08 25.6  &   618.0  &   1.27 \\
GMRTJ083859+424414 & 08 38 59.3  &  42 44 14.4  &    60.5  &   1.35 \\
GMRTJ083934+463534 & 08 39 34.3  &  46 35 34.0  &   924.1  &   1.21 \\
GMRTJ083957+432301 & 08 39 57.7  &  43 23 01.1  &   225.8  &   1.10 \\
GMRTJ084005+424657 & 08 40 05.8  &  42 46 57.8  &    24.2  &   1.20 \\
GMRTJ084012+432511 & 08 40 12.3  &  43 25 11.8  &    61.9  &   1.07 \\
GMRTJ084013+432319 & 08 40 13.5  &  43 23 19.8  &    66.8  &   1.20 \\
GMRTJ084021+431928 & 08 40 21.9  &  43 19 28.4  &    62.1  &   1.16 \\
GMRTJ084036+443132 & 08 40 36.6  &  44 31 32.2  &   132.7  &   1.02 \\
GMRTJ084045+460817 & 08 40 45.0  &  46 08 17.2  &    45.5  &   1.10 \\
GMRTJ084046+460719 & 08 40 46.3  &  46 07 19.9  &   197.1  &   1.01 \\
GMRTJ084105+445045 & 08 41 05.7  &  44 50 45.8  &   855.8  &   1.14 \\
GMRTJ084157+434105 & 08 41 57.5  &  43 41 05.6  &    86.2  &   1.02 \\
GMRTJ084211+445404 & 08 42 11.7  &  44 54 04.8  &   147.4  &   1.21 \\
GMRTJ084213+440637 & 08 42 13.8  &  44 06 37.7  &     9.2  &   1.00 \\
GMRTJ084220+424651 & 08 42 20.3  &  42 46 51.6  &   158.4  &   1.07 \\
GMRTJ084223+425532 & 08 42 23.8  &  42 55 32.1  &   100.0  &   1.19 \\
GMRTJ084242+453312 & 08 42 42.3  &  45 33 12.8  &    80.1  &   0.99 \\
GMRTJ084251+464505 & 08 42 51.5  &  46 45 05.1  &   198.5  &   1.04 \\
GMRTJ084300+453423 & 08 43 00.7  &  45 34 23.8  &   234.3  &   1.15 \\
GMRTJ084306+445552 & 08 43 06.8  &  44 55 52.1  &   191.7  &   1.16 \\
GMRTJ084311+433104 & 08 43 11.0  &  43 31 04.7  &   118.3  &   1.06 \\
GMRTJ084315+435630 & 08 43 15.8  &  43 56 30.9  &   124.5  &   1.00 \\
GMRTJ084322+461433 & 08 43 22.0  &  46 14 33.0  &   169.2  &   1.08 \\
GMRTJ084332+461703 & 08 43 32.4  &  46 17 03.3  &   192.8  &   1.03 \\
GMRTJ084402+462616 & 08 44 02.0  &  46 26 16.5  &    62.6  &   1.01 \\
GMRTJ084412+451559 & 08 44 12.4  &  45 15 59.7  &    38.1  &   1.04 \\
GMRTJ084415+430136 & 08 44 15.8  &  43 01 36.3  &    49.2  &   1.31 \\
GMRTJ084420+445802 & 08 44 20.4  &  44 58 02.3  &    29.0  &   0.99 \\
GMRTJ084437+442558 & 08 44 37.2  &  44 25 58.5  &    84.0  &   1.53 \\
GMRTJ084505+442543 & 08 45 05.5  &  44 25 43.4  &    28.0  &   1.05 \\
GMRTJ084515+431304 & 08 45 15.1  &  43 13 04.7  &   129.1  &   0.99 \\
GMRTJ084519+434953 & 08 45 19.9  &  43 49 53.1  &   342.7  &   1.04 \\
GMRTJ084522+434350 & 08 45 22.6  &  43 43 50.0  &   219.9  &   0.99 \\
GMRTJ084527+465014 & 08 45 27.6  &  46 50 14.6  &  1439.6  &   1.06 \\
GMRTJ084528+440732 & 08 45 28.7  &  44 07 32.4  &   141.8  &   1.48 \\
\end{tabular}
\end{table}

\begin{table}
\noindent {\bf Table 4. Contd}\\
\begin{tabular}{lllrl}
                       &               &                &          &        \\
NAME                   &  RA           &  DEC           & S$_{150}$ & $\alpha$ \\
                       &               &                &          &        \\
GMRTJ084533+455835 & 08 45 33.2  &  45 58 35.0  &    23.9  &   1.06 \\
GMRTJ084542+461535 & 08 45 42.9  &  46 15 35.9  &   753.3  &   1.11 \\
GMRTJ084559+425705 & 08 45 59.6  &  42 57 05.9  &   191.1  &   1.06 \\
GMRTJ084616+463234 & 08 46 16.2  &  46 32 34.3  &   150.3  &   0.99 \\
GMRTJ084627+453128 & 08 46 27.2  &  45 31 28.7  &   233.9  &   1.15 \\
GMRTJ084635+455030 & 08 46 35.5  &  45 50 30.2  &    19.4  &   1.01 \\
GMRTJ084644+432903 & 08 46 44.7  &  43 29 03.1  &   371.8  &   1.06 \\
GMRTJ084654+455935 & 08 46 54.4  &  45 59 35.7  &   219.7  &   1.11 \\
GMRTJ084711+454015 & 08 47 11.5  &  45 40 15.9  &  1704.3  &   1.04 \\
GMRTJ084728+430551 & 08 47 28.1  &  43 05 51.9  &   124.0  &   1.09 \\
GMRTJ084730+440822 & 08 47 30.5  &  44 08 22.5  &    63.6  &   1.11 \\
GMRTJ084737+461405 & 08 47 37.5  &  46 14 05.2  &  2819.6  &   1.16 \\
GMRTJ084742+451726 & 08 47 42.4  &  45 17 26.8  &    11.0  &   1.07 \\
GMRTJ084743+431711 & 08 47 43.7  &  43 17 11.2  &    17.4  &   1.01 \\
GMRTJ084747+451016 & 08 47 47.7  &  45 10 16.7  &   196.9  &   1.06 \\
GMRTJ084748+441228 & 08 47 48.2  &  44 12 28.7  &    50.3  &   1.08 \\
GMRTJ084755+445359 & 08 47 55.7  &  44 53 59.3  &   143.0  &   1.07 \\
GMRTJ084804+425728 & 08 48 04.3  &  42 57 28.0  &    90.4  &   1.16 \\
GMRTJ084811+460025 & 08 48 11.7  &  46 00 25.3  &    37.5  &   1.61 \\
GMRTJ084824+432240 & 08 48 24.5  &  43 22 40.2  &   132.3  &   1.23 \\
GMRTJ084842+455357 & 08 48 42.3  &  45 53 57.4  &   175.5  &   1.04 \\
GMRTJ084901+445049 & 08 49 01.2  &  44 50 49.3  &    18.8  &   1.88 \\
GMRTJ084938+442311 & 08 49 38.6  &  44 23 11.5  &   686.3  &   1.09 \\
GMRTJ084943+453750 & 08 49 43.1  &  45 37 50.5  &    12.8  &   1.28 \\
GMRTJ085018+432248 & 08 50 18.2  &  43 22 48.3  &   153.7  &   1.19 \\
GMRTJ085019+451435 & 08 50 19.4  &  45 14 35.1  &    18.7  &   1.34 \\
GMRTJ085028+445522 & 08 50 28.3  &  44 55 22.3  &   108.6  &   1.07 \\
GMRTJ085038+445637 & 08 50 38.0  &  44 56 37.5  &    41.9  &   1.45 \\
GMRTJ085044+454559 & 08 50 44.6  &  45 45 59.0  &    37.5  &   1.11 \\
GMRTJ085052+460440 & 08 50 52.9  &  46 04 40.0  &   110.6  &   1.27 \\
GMRTJ085132+441952 & 08 51 32.1  &  44 19 52.4  &   774.6  &   1.15 \\
GMRTJ085157+453448 & 08 51 57.9  &  45 34 48.8  &    66.5  &   1.08 \\
GMRTJ085202+435054 & 08 52 02.6  &  43 50 54.4  &    24.1  &   1.13 \\
GMRTJ085308+441109 & 08 53 08.3  &  44 11 09.0  &    92.6  &   1.09 \\
GMRTJ085322+450357 & 08 53 22.3  &  45 03 57.3  &   907.7  &   1.01 \\
GMRTJ085330+452256 & 08 53 30.3  &  45 22 56.5  &   162.1  &   1.13 \\
GMRTJ085406+445052 & 08 54 06.3  &  44 50 52.5  &    17.1  &   1.09 \\
\end{tabular}
\end{table}

\subsection{Sample of steep spectrum sources}

We present the sample of steep spectrum radio sources which do not have SDSS
counterparts in Table 4. 
The median 150 MHz flux density for these sources is $\sim$ 100 mJy, which is
more than an order of magnitude fainter than the median flux density at 150
MHz for known HzRGs (section 2).  If we further divide the sample into
two parts, one with 1 $< \alpha < 1.3$ and $\alpha > 1.3$, we find
that the fraction of sources without identification in SDSS is 63\%
for sources with 1 $< \alpha < 1.3$ and 68\% for sources with $\alpha
> 1.3$.  Although this difference is not statistically significant, the
trend of increased fraction of sources without identification in SDSS
for sources with steeper spectra goes in the expected direction.

A few examples of the radio spectra among these sources are 
given in Figure  7. One of the sources with
steepest spectrum ($\alpha$ = 1.53; GMRT084437+442558) has measurements at 150, 327, 1412 and 4860 MHz.
We have inspected the spectra where three or more frequency measurements 
are available for signatures of spectral curvature. There is no clear indication
of spectral curvature in most of the sources. Since the frequency range is 
just one order of magnitude, higher frequency radio data are needed to 
further investigate the question of spectral curvature. The absence of spectral curvature in these
data is consistent with the results of Klamer et al (2006), also
argued that the spectral curvature is unlikely to be the 
reason for steepening of the radio spectra at high-redshifts.

One of the steep spectrum source without SDSS counterpart (GMRTJ084533+455835),
is unresolved at 150 MHz, but resolves into a clear compact FRII source
of about 8 arcsec size in VLA FIRST. Using the 150 MHz flux density and FRI/FRII
break luminosity, we estimate that this source should be at a redshift of $\sim$
2 or higher (for its luminosity to be above the FRI/FRII break).

\subsection{Notes on individual sources:}

Among the 98 steep spectrum sources that do not have counterparts in
SDSS, 68 are unresolved sources and have only one component in 1.4 GHz FIRST as
well. Among the remaining 30 sources, we discuss those  which show
significant structure, many of which are unlikely to be at high-redshift (Figure 9),
but remained unidentified in SDSS either due to confusion in getting
the correct radio component and position or due to dust obscuration
in the host galaxy.

\noindent {\bf GMRTJ083349+432020:} This is a well defined double
radio source at 150 MHz as well as in FIRST. The lobes appear to be of
FR-II type, however they are not compact, and seem to have 'diffused
out'. Since the core is not visible, the SDSS identification is
difficult. The spectral index of this source is 1.37. From the relaxed
morphology of the lobes, this could be an example of 'dead AGN', as
highlighted recently using 74 MHz and NVSS data (Dwarakanath \& Kale,
2009).

\noindent {\bf GMRTJ083405+453443:} The 150 MHz map does not reveal any
structure. However, the FIRST image shows interesting 'one-sided jet'
morphology. There is a compact component appearing like a core at the north-east end
and extended emission towards the south-west resembling a lobe, but lacks clear hotspot near the edge.
The plume like structure to the right is also unusual. This is an interesting
source worth further high-resolution observations to understand the unusual morphology.

\noindent {\bf GMRTJ083507+453628:} The structure at 150 MHz just
resolves into two lobes, which are mostly resolved out in FIRST. This
indicates that the lobes do not have  compact hotspots. This
steep spectrum is suggestive of the lobes belonging to 'dead AGN'.

\noindent {\bf GMRTJ083833+461651:} This is one of the steepest spectrum source in the 
sample with the spectral index of 1.53. The 150 MHz images does not resolve two lobes
clearly. The FIRST image shows three structures. However, the central emitting region
does not appear like a core. It is possible that the two emitting regions on either
side are a pair of lobes, and the central component is an unrelated source.

\noindent {\bf GMRTJ084105+445045:} An asymmetric triple with a bright core at 150 MHz?
The FIRST image resolves this into a linear double and a likely unrelated source.
At first sight, this double radio source resembles the 'double-double' morphology,
however closer inspection reveals that the inner pair of lobes is more relaxed than
the outer pair, which is inconsistent with the standard double-double 
morphology (Schoenmakers et al. 2000).

\noindent {\bf GMRTJ084711+454015:} Appears to be a double at 150 MHz. 
The lobes on either side showed extended diffuse emission.
The FIRST image clearly resolves the eastern lobe into a diffuse
'relic' type morphology and the west lobe into a  head-tail looking structure.
There is no known cluster in this region. The spectral index of diffuse region alone is 1.2

\section{Concluding Remarks}

The radio spectral-index redshift correlation is the most efficient
method used to detect high-redshift radio galaxies. Up to now, about 45
radio galaxies beyond redshift of 3 have been discovered using this
method. We have shown that this known high-z population is just the tip of the
iceberg, in the sense that they are the most luminous objects in this
class.  Radio sources which are 10 to 100 times less luminous than these are yet
to be discovered, and these are essential to fully understand the cosmological evolution
of radio galaxies.  We have initiated a major
programme to search for steep spectrum radio sources with the GMRT with the
aim to detect high-redshift radio galaxies of moderate luminosity.
The fields for this programme are carefully chosen such that extensive 
data already exist at higher radio frequencies and deep optical imaging
and/or spectroscopy is also available for most of the fields.

Here we have presented the results from deep 150 MHz low frequency
radio observations with the Giant Metrewave Radio Telescope (GMRT), India
reaching an rms noise of of $\sim$ 0.7 mJy/beam and a  resolution of $\sim$
20$^{''}$. Further, several deep observations exist, mainly at 327, 610 and
1412 MHz for this field. The radio spectral index analysis of the
sources in the field was done using these deep observations and also
using the WENSS at 325 MHz and the NVSS and FIRST at 1400 MHz.  We have
demonstrated that this GMRT survey can search for high-redshift radio
galaxies more than an order of magnitude fainter in luminosity
compared to most of the known HzRGs.  We provide a sample of about 100
candidate HzRGs with spectral index steeper than 1 and no optical
counterpart to the SDSS sensitivity limits. A significant fraction of the sources are
compact, and are strong candidates for high-redshift radio galaxies.  
These sources will need to be followed up at optical and
near-infrared bands to estimate their redshifts.

\section*{Acknowledgments }

We thank the referee for insightful suggestions and Gopal-Krishna for comments on the manuscript.
GMRT is operated by the National Centre for Radio Astrophysics of the Tata
Institute of Fundamental Research. This research has made use of the
NASA/IPAC Extragalactic Database (NED) which is operated by the Jet
Propulsion Laboratory, California Institute of Technology, under
contract with the National Aeronautics and Space Administration.
Funding for the SDSS and SDSS-II has been provided by the Alfred
P. Sloan Foundation, the Participating Institutions, the National
Science Foundation, the U.S. Department of Energy, the National
Aeronautics and Space Administration, the Japanese Monbukagakusho, the
Max Planck Society, and the Higher Education Funding Council for
England. The SDSS Web Site is http://www.sdss.org/.
The SDSS is managed by the Astrophysical Research Consortium for the
Participating Institutions. The Participating Institutions are the
American Museum of Natural History, Astrophysical Institute Potsdam,
University of Basel, University of Cambridge, Case Western Reserve
University, University of Chicago, Drexel University, Fermilab, the
Institute for Advanced Study, the Japan Participation Group, Johns
Hopkins University, the Joint Institute for Nuclear Astrophysics, the
Kavli Institute for Particle Astrophysics and Cosmology, the Korean
Scientist Group, the Chinese Academy of Sciences (LAMOST), Los Alamos
National Laboratory, the Max-Planck-Institute for Astronomy (MPIA),
the Max-Planck-Institute for Astrophysics (MPA), New Mexico State
University, Ohio State University, University of Pittsburgh,
University of Portsmouth, Princeton University, the United States
Naval Observatory, and the University of Washington.

%\end{document}
  
\newpage

%
% THIS TABLE IS FOR ONLINE VERSION ONLY
%

\begin{table*}
\caption{Table of GMRT 150 MHz sources (see also Table 3).
Column 1: Source Name; Columns 2-4: RA; Columns 5-7: DEC; Columns
8-17: The flux density at 150 MHz(GMRT), 327 MHz(WSRT), 327 MHz(WENSS), 610 MHz(WSRT),
1412 MHz(WSRT, Windhorst et al. 1984), 1412 MHz(WSRT, Oort et al. 1987), 1400 MHz(NVSS), 1400 MHz(FIRST), 1462 MHz(VLA) and 4860 MHz(VLA) respectively;
Column 18: Spectral index from the
blind straightline fit, which may not be valid for a small fraction ($\sim$ 3\%) of sources.  }
\scriptsize
\begin{tabular}{@{\extracolsep{-5pt}}llllllllllllllllll}
NAME               & \multicolumn{3}{c}{RA (J2000)}  & \multicolumn{3}{c}{DEC (J2000)}   & S$_{150}$ & S$_{327}$   & S$_{WENSS}$      & S$_{610}$   & S$_{1412^a}$   & S$_{1412^b}$   &  S$_{NVSS}$     &  S$_{FIRST}$     & S$_{1462}$   &  S$_{4860}$   & $\alpha$\\
                   & hh & mm & ss.s & dd  & mm & ss.s & mJy     & mJy & mJy    &mJy  & mJy & mJy & mJy    &  mJy   & mJy & mJy  &      \\
                   &    &    &      &     &    &      &         &     &        &     &     &     &        &        &     &      &      \\
GMRTJ083035+454329 & 08 & 30 & 35.7 &  45 & 43 & 29.6 & 11342.3 & $-$ & 3049.0 & $-$ & $-$ & $-$ &  614.4 &  608.3 & $-$ & $-$ & 1.48 \\
GMRTJ083752+445024 & 08 & 37 & 52.9 &  44 & 50 & 24.6 & 11158.2 & 4561.5 & 4662.0 & 2890.0 & 1519.2 & $-$ & 1528.9 & 1139.8 & $-$ & $-$ & 1.01 \\
GMRTJ084639+462141 & 08 & 46 & 39.7 &  46 & 21 & 41.9 &  3201.8 & 1299.4 & 1303.0 & $-$ & $-$ & $-$ &  362.3 &  380.0 & $-$ & $-$ & 1.03 \\
GMRTJ084737+461405 & 08 & 47 & 37.6 &  46 & 14 &  5.3 &  2819.6 & 1028.8 & 1043.0 & $-$ & $-$ & $-$ &  262.1 &  217.3 & $-$ & $-$ & 1.16 \\
GMRTJ084034+465112 & 08 & 40 & 34.3 &  46 & 51 & 12.8 &  2225.0 & $-$ &  928.0 & $-$ & $-$ & $-$ &  287.8 &  297.6 & $-$ & $-$ & 0.94 \\
\end{tabular} 
\\
$^a$ Windhorst et al. 1984 ~~ 
$^b$ Oort et al. 1987
\end{table*}
\begin{table*}
\begin{tabular}{@{\extracolsep{-5pt}}llllllllllllllllll}
NAME               & \multicolumn{3}{c}{RA (J2000)}  & \multicolumn{3}{c}{DEC (J2000)}   & S$_{150}$ & S$_{327}$   & S$_{WENSS}$      & S$_{610}$   & S$_{1412^a}$   & S$_{1412^b}$   &  S$_{NVSS}$     &  S$_{FIRST}$     & S$_{1462}$   &  S$_{4860}$   & $\alpha$\\
                   & hh & mm & ss.s & dd  & mm & ss.s & mJy     & mJy & mJy    &mJy  & mJy & mJy & mJy    &  mJy   & mJy & mJy  &      \\
                   &    &    &      &     &    &      &         &     &        &     &     &     &        &        &     &      &      \\
GMRTJ084300+453423 & 08 & 43 &  0.8 &  45 & 34 & 23.8 &   234.3 &   83.2 &   78.0 &   42.0 & $-$ & $-$ &   21.5 &   15.6 & $-$ & $-$ & 1.15 \\
GMRTJ084627+453128 & 08 & 46 & 27.2 &  45 & 31 & 28.7 &   233.9 &   89.4 &  102.0 &   46.1 & $-$ & $-$ &   17.1 &   19.0 & $-$ & $-$ & 1.15 \\
GMRTJ083957+432301 & 08 & 39 & 57.8 &  43 & 23 &  1.2 &   225.8 &   57.9 &   64.0 & $-$ & $-$ & $-$ &   21.4 &   18.2 & $-$ & $-$ & 1.10 \\
GMRTJ084522+434350 & 08 & 45 & 22.6 &  43 & 43 & 50.1 &   219.9 &   81.6 &   76.0 &   47.3 & $-$ & $-$ &   24.1 &   24.6 & $-$ & $-$ & 0.99 \\
GMRTJ084654+455935 & 08 & 46 & 54.4 &  45 & 59 & 35.7 &   219.7 &   79.7 &   83.0 & $-$ & $-$ & $-$ &   17.5 &   19.9 & $-$ & $-$ & 1.11 \\
\end{tabular}
\end{table*}
\begin{table*}
\begin{tabular}{@{\extracolsep{-5pt}}llllllllllllllllll}
NAME               & \multicolumn{3}{c}{RA (J2000)}  & \multicolumn{3}{c}{DEC (J2000)}   & S$_{150}$ & S$_{327}$   & S$_{WENSS}$      & S$_{610}$   & S$_{1412^a}$   & S$_{1412^b}$   &  S$_{NVSS}$     &  S$_{FIRST}$     & S$_{1462}$   &  S$_{4860}$   & $\alpha$\\
                   & hh & mm & ss.s & dd  & mm & ss.s & mJy     & mJy & mJy    &mJy  & mJy & mJy & mJy    &  mJy   & mJy & mJy  &      \\
                   &    &    &      &     &    &      &         &     &        &     &     &     &        &        &     &      &      \\
GMRTJ084515+431304 & 08 & 45 & 15.1 &  43 & 13 &  4.7 &   129.1 &   38.5 &   41.0 & $-$ & $-$ & $-$ &   12.9 &   15.1 & $-$ & $-$ & 0.99 \\
GMRTJ083348+441303 & 08 & 33 & 48.5 &  44 & 13 &  3.5 &   128.4 &   47.6 &   46.0 & $-$ & $-$ & $-$ &   16.1 &   15.9 & $-$ & $-$ & 0.93 \\
GMRTJ083638+440516 & 08 & 36 & 38.2 &  44 &  5 & 16.0 &   126.9 &   65.8 &   71.0 &   48.0 & $-$ & $-$ &   28.8 &   25.8 & $-$ & $-$ & 0.68 \\
GMRTJ084745+433813 & 08 & 47 & 45.4 &  43 & 38 & 13.8 &   126.6 &   52.1 &   56.0 &   42.0 & $-$ & $-$ &   23.7 &   22.4 & $-$ & $-$ & 0.74 \\
GMRTJ084215+452536 & 08 & 42 & 15.6 &  45 & 25 & 36.1 &   126.3 &   80.6 &   78.0 &   61.2 &   75.6 & $-$ &   74.0 &   71.6 &  304.0 & $-$ & -0.09 \\
\end{tabular}
\end{table*}
\begin{table*}
\begin{tabular}{@{\extracolsep{-5pt}}llllllllllllllllll}
NAME               & \multicolumn{3}{c}{RA (J2000)}  & \multicolumn{3}{c}{DEC (J2000)}   & S$_{150}$ & S$_{327}$   & S$_{WENSS}$      & S$_{610}$   & S$_{1412^a}$   & S$_{1412^b}$   &  S$_{NVSS}$     &  S$_{FIRST}$     & S$_{1462}$   &  S$_{4860}$   & $\alpha$\\
                   & hh & mm & ss.s & dd  & mm & ss.s & mJy     & mJy & mJy    &mJy  & mJy & mJy & mJy    &  mJy   & mJy & mJy  &      \\
                   &    &    &      &     &    &      &         &     &        &     &     &     &        &        &     &      &      \\
GMRTJ084541+444011 & 08 & 45 & 41.3 &  44 & 40 & 11.0 &    70.6 &   32.9 &   34.0 &   21.5 &   11.7 & $-$ &   11.1 &   12.6 &   11.8 & $-$ & 0.78 \\
GMRTJ085001+462600 & 08 & 50 &  1.3 &  46 & 26 &  0.7 &    70.1 &   55.1 &   54.0 & $-$ & $-$ & $-$ &   16.0 &   21.3 & $-$ & $-$ & 0.61 \\
GMRTJ084943+435806 & 08 & 49 & 43.8 &  43 & 58 &  6.5 &    69.6 &   32.6 &   14.0 &   23.8 &   12.4 & $-$ &   13.4 &   12.6 & $-$ & $-$ & 0.69 \\
GMRTJ084757+441210 & 08 & 47 & 57.0 &  44 & 12 & 10.8 &    69.0 & $-$ &   22.0 &   19.5 &   10.3 & $-$ &   10.6 &   10.7 & $-$ & $-$ & 0.81 \\
GMRTJ083637+431808 & 08 & 36 & 37.2 &  43 & 18 &  8.2 &    68.6 & $-$ &   20.0 & $-$ & $-$ & $-$ &   10.2 &   11.1 & $-$ & $-$ & 0.81 \\
\end{tabular}
\end{table*}
\begin{table*}
\begin{tabular}{@{\extracolsep{-5pt}}llllllllllllllllll}
NAME               & \multicolumn{3}{c}{RA (J2000)}  & \multicolumn{3}{c}{DEC (J2000)}   & S$_{150}$ & S$_{327}$   & S$_{WENSS}$      & S$_{610}$   & S$_{1412^a}$   & S$_{1412^b}$   &  S$_{NVSS}$     &  S$_{FIRST}$     & S$_{1462}$   &  S$_{4860}$   & $\alpha$\\
                   & hh & mm & ss.s & dd  & mm & ss.s & mJy     & mJy & mJy    &mJy  & mJy & mJy & mJy    &  mJy   & mJy & mJy  &      \\
                   &    &    &      &     &    &      &         &     &        &     &     &     &        &        &     &      &      \\
GMRTJ084045+460817 & 08 & 40 & 45.0 &  46 &  8 & 17.2 &    45.5 & $-$ & $-$ & $-$ & $-$ & $-$ &    3.1 &    4.8 & $-$ & $-$ & 1.10 \\
GMRTJ083658+460845 & 08 & 36 & 58.7 &  46 &  8 & 45.6 &    45.0 &   23.4 &   20.0 & $-$ & $-$ & $-$ &    9.7 &    5.1 & $-$ & $-$ & 0.81 \\
GMRTJ083018+453459 & 08 & 30 & 18.8 &  45 & 34 & 59.0 &    44.7 & $-$ &   30.0 & $-$ & $-$ & $-$ &    9.8 &   10.0 & $-$ & $-$ & 0.68 \\
GMRTJ084643+462149 & 08 & 46 & 43.6 &  46 & 21 & 49.5 &    44.4 & $-$ & $-$ & $-$ & $-$ & $-$ & $-$ & $-$ & $-$ & $-$ & $-$ \\
GMRTJ084837+442936 & 08 & 48 & 37.0 &  44 & 29 & 36.9 &    44.2 &   27.3 &   19.0 &   17.7 &   11.0 &   12.0 &   12.3 &   11.5 & $-$ & $-$ & 0.56 \\
\end{tabular}
\end{table*}
\begin{table*}
\begin{tabular}{@{\extracolsep{-5pt}}llllllllllllllllll}
NAME               & \multicolumn{3}{c}{RA (J2000)}  & \multicolumn{3}{c}{DEC (J2000)}   & S$_{150}$ & S$_{327}$   & S$_{WENSS}$      & S$_{610}$   & S$_{1412^a}$   & S$_{1412^b}$   &  S$_{NVSS}$     &  S$_{FIRST}$     & S$_{1462}$   &  S$_{4860}$   & $\alpha$\\
                   & hh & mm & ss.s & dd  & mm & ss.s & mJy     & mJy & mJy    &mJy  & mJy & mJy & mJy    &  mJy   & mJy & mJy  &      \\
                   &    &    &      &     &    &      &         &     &        &     &     &     &        &        &     &      &      \\
GMRTJ084647+451848 & 08 & 46 & 47.0 &  45 & 18 & 48.5 &    30.2 &   12.6 & $-$ &    9.5 & $-$ &    2.6 &    4.3 &    3.3 & $-$ & $-$ & 0.96 \\
GMRTJ084524+434917 & 08 & 45 & 24.8 &  43 & 49 & 17.6 &    29.9 & $-$ & $-$ &   21.5 & $-$ & $-$ &   14.1 &   14.3 & $-$ & $-$ & 0.33 \\
GMRTJ084201+440552 & 08 & 42 &  1.6 &  44 &  5 & 52.8 &    29.5 &   17.7 & $-$ &   23.5 & $-$ & $-$ &   20.3 &   18.3 &   22.5 & $-$ & 0.12 \\
GMRTJ085311+440913 & 08 & 53 & 11.6 &  44 &  9 & 13.2 &    29.4 & $-$ & $-$ & $-$ & $-$ & $-$ & $-$ & $-$ & $-$ & $-$ & $-$ \\
GMRTJ084420+445802 & 08 & 44 & 20.5 &  44 & 58 &  2.3 &    29.0 &   15.6 & $-$ &    5.9 &    1.9 & $-$ &    3.4 &    3.3 &    4.7 & $-$ & 0.99 \\
\end{tabular}
\end{table*}
\begin{table*}
\begin{tabular}{@{\extracolsep{-5pt}}llllllllllllllllll}
NAME               & \multicolumn{3}{c}{RA (J2000)}  & \multicolumn{3}{c}{DEC (J2000)}   & S$_{150}$ & S$_{327}$   & S$_{WENSS}$      & S$_{610}$   & S$_{1412^a}$   & S$_{1412^b}$   &  S$_{NVSS}$     &  S$_{FIRST}$     & S$_{1462}$   &  S$_{4860}$   & $\alpha$\\
                   & hh & mm & ss.s & dd  & mm & ss.s & mJy     & mJy & mJy    &mJy  & mJy & mJy & mJy    &  mJy   & mJy & mJy  &      \\
                   &    &    &      &     &    &      &         &     &        &     &     &     &        &        &     &      &      \\
GMRTJ083451+443811 & 08 & 34 & 51.9 &  44 & 38 & 11.8 &    21.8 & $-$ & $-$ &   12.1 & $-$ & $-$ &    5.1 &    5.8 & $-$ & $-$ & 0.60 \\
GMRTJ084857+454946 & 08 & 48 & 57.5 &  45 & 49 & 46.1 &    21.7 &   24.9 & $-$ & $-$ & $-$ & $-$ &   12.1 &   12.3 & $-$ & $-$ & 0.29 \\
GMRTJ084701+454052 & 08 & 47 &  1.8 &  45 & 40 & 52.5 &    21.5 & $-$ & $-$ & $-$ & $-$ & $-$ & $-$ &    1.0 & $-$ & $-$ & 1.38 \\
GMRTJ084905+454132 & 08 & 49 &  5.6 &  45 & 41 & 32.0 &    21.5 & $-$ & $-$ & $-$ & $-$ & $-$ &    5.5 &    4.9 & $-$ & $-$ & 0.64 \\
GMRTJ084148+444238 & 08 & 41 & 48.9 &  44 & 42 & 38.3 &    21.5 &    9.4 & $-$ &    7.0 & $-$ & $-$ & $-$ & $-$ & $-$ & $-$ & 0.84 \\
\end{tabular}
\end{table*}
\begin{table*}
\begin{tabular}{@{\extracolsep{-5pt}}llllllllllllllllll}
NAME               & \multicolumn{3}{c}{RA (J2000)}  & \multicolumn{3}{c}{DEC (J2000)}   & S$_{150}$ & S$_{327}$   & S$_{WENSS}$      & S$_{610}$   & S$_{1412^a}$   & S$_{1412^b}$   &  S$_{NVSS}$     &  S$_{FIRST}$     & S$_{1462}$   &  S$_{4860}$   & $\alpha$\\
                   & hh & mm & ss.s & dd  & mm & ss.s & mJy     & mJy & mJy    &mJy  & mJy & mJy & mJy    &  mJy   & mJy & mJy  &      \\
                   &    &    &      &     &    &      &         &     &        &     &     &     &        &        &     &      &      \\
GMRTJ083552+452857 & 08 & 35 & 52.7 &  45 & 28 & 57.7 &    16.0 & $-$ & $-$ & $-$ & $-$ & $-$ &    3.5 &    2.7 & $-$ & $-$ & 0.73 \\
GMRTJ083548+440821 & 08 & 35 & 48.8 &  44 &  8 & 21.9 &    15.9 & $-$ & $-$ & $-$ & $-$ & $-$ &    3.1 &    2.3 & $-$ & $-$ & 0.80 \\
GMRTJ084448+455714 & 08 & 44 & 48.5 &  45 & 57 & 14.0 &    15.9 & $-$ & $-$ & $-$ & $-$ & $-$ &    2.7 &    1.8 & $-$ & $-$ & 0.88 \\
GMRTJ083928+453005 & 08 & 39 & 28.9 &  45 & 30 &  5.4 &    15.8 &    8.8 & $-$ &    6.0 & $-$ & $-$ &    3.4 &    2.8 & $-$ & $-$ & 0.73 \\
GMRTJ084111+444232 & 08 & 41 & 11.2 &  44 & 42 & 32.2 &    15.8 &   12.2 & $-$ &    9.7 &    4.8 & $-$ &    4.7 &    4.5 & $-$ & $-$ & 0.56 \\
\end{tabular}
\end{table*}
\begin{table*}
\begin{tabular}{@{\extracolsep{-5pt}}llllllllllllllllll}
NAME               & \multicolumn{3}{c}{RA (J2000)}  & \multicolumn{3}{c}{DEC (J2000)}   & S$_{150}$ & S$_{327}$   & S$_{WENSS}$      & S$_{610}$   & S$_{1412^a}$   & S$_{1412^b}$   &  S$_{NVSS}$     &  S$_{FIRST}$     & S$_{1462}$   &  S$_{4860}$   & $\alpha$\\
                   & hh & mm & ss.s & dd  & mm & ss.s & mJy     & mJy & mJy    &mJy  & mJy & mJy & mJy    &  mJy   & mJy & mJy  &      \\
                   &    &    &      &     &    &      &         &     &        &     &     &     &        &        &     &      &      \\
GMRTJ083453+431144 & 08 & 34 & 53.6 &  43 & 11 & 44.6 &    12.6 & $-$ & $-$ & $-$ & $-$ & $-$ &    2.9 &    2.7 & $-$ & $-$ & 0.67 \\
GMRTJ083632+445510 & 08 & 36 & 32.0 &  44 & 55 & 10.3 &    12.6 &    6.0 & $-$ &    5.0 &    3.1 & $-$ &    3.3 &    2.2 & $-$ & $-$ & 0.64 \\
GMRTJ083701+435554 & 08 & 37 &  1.9 &  43 & 55 & 54.6 &    12.6 & $-$ & $-$ & $-$ & $-$ & $-$ & $-$ &    1.0 & $-$ & $-$ & 1.13 \\
GMRTJ084427+444305 & 08 & 44 & 27.4 &  44 & 43 &  5.5 &    12.6 &    6.1 & $-$ &    4.4 &    1.3 & $-$ &    3.7 &    1.7 &    1.9 &   0.54 & 0.84 \\
GMRTJ084542+452255 & 08 & 45 & 42.4 &  45 & 22 & 55.4 &    12.4 & $-$ & $-$ & $-$ & $-$ & $-$ & $-$ &    2.6 & $-$ & $-$ & 0.70 \\
\end{tabular}
\end{table*}
\begin{table*}
\begin{tabular}{@{\extracolsep{-5pt}}llllllllllllllllll}
NAME               & \multicolumn{3}{c}{RA (J2000)}  & \multicolumn{3}{c}{DEC (J2000)}   & S$_{150}$ & S$_{327}$   & S$_{WENSS}$      & S$_{610}$   & S$_{1412^a}$   & S$_{1412^b}$   &  S$_{NVSS}$     &  S$_{FIRST}$     & S$_{1462}$   &  S$_{4860}$   & $\alpha$\\
                   & hh & mm & ss.s & dd  & mm & ss.s & mJy     & mJy & mJy    &mJy  & mJy & mJy & mJy    &  mJy   & mJy & mJy  &      \\
                   &    &    &      &     &    &      &         &     &        &     &     &     &        &        &     &      &      \\
GMRTJ083942+453007 & 08 & 39 & 42.7 &  45 & 30 &  7.7 &     9.7 & $-$ & $-$ &    6.4 & $-$ & $-$ &    4.3 &    5.0 & $-$ & $-$ & 0.33 \\
GMRTJ084347+440452 & 08 & 43 & 47.7 &  44 &  4 & 52.6 &     9.7 & $-$ & $-$ & $-$ & $-$ & $-$ & $-$ & $-$ & $-$ & $-$ & $-$ \\
GMRTJ084246+431017 & 08 & 42 & 46.7 &  43 & 10 & 17.6 &     9.7 & $-$ & $-$ & $-$ & $-$ & $-$ &    3.5 &    1.9 & $-$ & $-$ & 0.58 \\
GMRTJ084454+444620 & 08 & 44 & 54.5 &  44 & 46 & 20.1 &     9.7 &    4.9 & $-$ &    2.5 &    1.7 & $-$ & $-$ &    1.1 &    1.8 &   0.74 & 0.77 \\
GMRTJ083512+433301 & 08 & 35 & 12.1 &  43 & 33 &  1.5 &     9.6 & $-$ & $-$ & $-$ & $-$ & $-$ &    3.2 &    2.7 & $-$ & $-$ & 0.53 \\
\end{tabular}
\end{table*}
\begin{table*}
\begin{tabular}{@{\extracolsep{-5pt}}llllllllllllllllll}
NAME               & \multicolumn{3}{c}{RA (J2000)}  & \multicolumn{3}{c}{DEC (J2000)}   & S$_{150}$ & S$_{327}$   & S$_{WENSS}$      & S$_{610}$   & S$_{1412^a}$   & S$_{1412^b}$   &  S$_{NVSS}$     &  S$_{FIRST}$     & S$_{1462}$   &  S$_{4860}$   & $\alpha$\\
                   & hh & mm & ss.s & dd  & mm & ss.s & mJy     & mJy & mJy    &mJy  & mJy & mJy & mJy    &  mJy   & mJy & mJy  &      \\
                   &    &    &      &     &    &      &         &     &        &     &     &     &        &        &     &      &      \\
GMRTJ084326+433502 & 08 & 43 & 26.1 &  43 & 35 &  2.5 &     7.4 & $-$ & $-$ & $-$ & $-$ & $-$ & $-$ & $-$ & $-$ & $-$ & $-$ \\
GMRTJ084100+431705 & 08 & 41 &  0.6 &  43 & 17 &  5.7 &     7.3 & $-$ & $-$ & $-$ & $-$ & $-$ & $-$ & $-$ & $-$ & $-$ & $-$ \\
GMRTJ084030+443106 & 08 & 40 & 30.9 &  44 & 31 &  6.7 &     7.3 & $-$ & $-$ & $-$ & $-$ & $-$ & $-$ & $-$ & $-$ & $-$ & $-$ \\
GMRTJ084128+443337 & 08 & 41 & 28.8 &  44 & 33 & 37.9 &     7.2 &    4.0 & $-$ &    4.7 &    2.2 & $-$ & $-$ &    3.2 & $-$ & $-$ & 0.41 \\
GMRTJ084340+443922 & 08 & 43 & 40.9 &  44 & 39 & 22.1 &     7.2 &    3.4 & $-$ & $-$ &    1.0 & $-$ & $-$ &    0.8 &    1.3 & $-$ & 0.85 \\
\end{tabular}
\end{table*}
\begin{table*}
\begin{tabular}{@{\extracolsep{-5pt}}llllllllllllllllll}
NAME               & \multicolumn{3}{c}{RA (J2000)}  & \multicolumn{3}{c}{DEC (J2000)}   & S$_{150}$ & S$_{327}$   & S$_{WENSS}$      & S$_{610}$   & S$_{1412^a}$   & S$_{1412^b}$   &  S$_{NVSS}$     &  S$_{FIRST}$     & S$_{1462}$   &  S$_{4860}$   & $\alpha$\\
                   & hh & mm & ss.s & dd  & mm & ss.s & mJy     & mJy & mJy    &mJy  & mJy & mJy & mJy    &  mJy   & mJy & mJy  &      \\
                   &    &    &      &     &    &      &         &     &        &     &     &     &        &        &     &      &      \\
GMRTJ084409+444743 & 08 & 44 &  9.2 &  44 & 47 & 43.2 &     5.1 & $-$ & $-$ & $-$ & $-$ & $-$ & $-$ & $-$ &    0.5 &   0.72 & 0.66 \\
GMRTJ084628+441741 & 08 & 46 & 28.5 &  44 & 17 & 41.5 &     5.1 & $-$ & $-$ &    1.7 &    2.0 & $-$ & $-$ & $-$ & $-$ & $-$ & 0.48 \\
GMRTJ084802+452444 & 08 & 48 &  2.6 &  45 & 24 & 44.2 &     5.1 & $-$ & $-$ & $-$ & $-$ & $-$ & $-$ & $-$ & $-$ & $-$ & $-$ \\
GMRTJ084003+455152 & 08 & 40 &  3.6 &  45 & 51 & 52.5 &     5.1 & $-$ & $-$ & $-$ & $-$ & $-$ &    3.4 &    1.7 & $-$ & $-$ & 0.32 \\
GMRTJ084315+450212 & 08 & 43 & 15.0 &  45 &  2 & 12.6 &     5.0 & $-$ & $-$ & $-$ & $-$ & $-$ &    2.1 &    2.0 & $-$ & $-$ & 0.40 \\
\hline
\end{tabular}
%\centerline{----------------------------------------------------------}
\end{table*}
\end{document}